\documentclass[10pt,journal]{IEEEtran}

\usepackage{cite}

\ifCLASSINFOpdf
\else
  \usepackage[dvips]{graphicx}
  \DeclareGraphicsExtensions{.eps}
\fi
\usepackage[cmex10]{amsmath}
\usepackage{amsfonts,amssymb,amsthm,mdwmath}
\usepackage{array}

\ifCLASSOPTIONcompsoc
  \usepackage[caption=false,font=normalsize,labelfont=sf,textfont=sf]{subfig}
\else
  \usepackage[caption=false,font=footnotesize]{subfig}
\fi

\usepackage{stfloats}
\usepackage{url}
\usepackage{color}
\usepackage{acronym}

\begin{document}
\acrodef{PPP}[PPP]{Poisson Point Process}
\acrodef{CDF}[CDF]{Cummulative Distribution Function}
\acrodef{PDF}[PDF]{Probability Distribution Function}
\acrodef{PMF}[PMF]{Probability Mass Function}
\acrodef{PCF}[PCF]{Pair Correlation Function}
\acrodef{PGFL}[PGFL]{Probability Generating Functional}
\acrodef{MGF}[MGF]{Moment Generating Function}
\acrodef{RV}[RV]{Random Variable}
\acrodef{1D}[1D]{one-dimensional}
\acrodef{i.i.d.}[i.i.d.]{independent and identically distributed}

\title{Moments of Interference in Vehicular Networks with Hardcore Headway Distance}
\author{Konstantinos Koufos and Carl P. Dettmann 
\thanks{K.~Koufos and C.P.~Dettmann are with the School of Mathematics, University of Bristol, BS8 1TW, Bristol, UK. \{K.Koufos, Carl.Dettmann\}@bristol.ac.uk} \protect \\ 
\thanks{This work was supported by the EPSRC grant number 
EP/N002458/1 for the project Spatially Embedded Networks. All underlying data are provided in full within this paper.}}
\maketitle
\begin{abstract}
Interference statistics in vehicular networks have long been studied using the Poisson Point Process (PPP) for the locations of vehicles. In roads with few number of lanes and restricted overtaking, this model becomes unrealistic because it assumes that the vehicles can come arbitrarily close to each other. In this paper, we model the headway distance (the distance between the head of a vehicle and the head of its follower) equal to the sum of a constant hardcore distance and an exponentially distributed random variable. We study the mean, the variance and the skewness of interference at the origin with this deployment model. Even though the pair correlation function becomes complicated, we devise simple formulae to capture the impact of hardcore distance on the variance of interference in comparison with a PPP model of equal intensity. In addition, we study the extreme scenario where the interference originates from a lattice. We show how to relate the variance of interference due to a lattice to that of a PPP under Rayleigh fading. 
\end{abstract}

\section{Introduction}
\label{sec:Introduction}
Interference statistics in wireless networks with unknown locations of users have long been studied using stochastic geometry~\cite{Haenggi2009}. Due to its analytical tractability, the \ac{PPP} is the most commonly employed model. A \ac{PPP} with non-homogeneous intensity has been used to capture a variable intensity of users due to mobility~\cite{Gong2014,Koufos2018}. The distribution of interferers in cellular uplink with a single interferer per Voronoi cell has also been approximated by non-uniform \ac{PPP}~\cite{Haenggi2017a}. Superposition of independent \acp{PPP} of different intensities is applicable to heterogeneous cellular networks~\cite{Dhillon2012}. By definition, a \ac{PPP} assumes that two points (or users) can come arbitrarily close to each other. This assumption may not be accurate due to physical constraints and/or medium access control. In this regard, determinantal point processes have been used to describe the deployment of real-world macro base stations~\cite{Miyoshi2014, Dhillon2015}, and Mat{\`e}rn point processes to model the locations of active transmitters in carrier sensing multiple access wireless ad hoc networks~\cite{Busson2009, Haenggi2011}. The point processes suggested in~\cite{Miyoshi2014, Dhillon2015,Busson2009, Haenggi2011,Gong2014,Koufos2018,Haenggi2017a,Dhillon2012} were constructed to study either planar cellular networks or \ac{1D} ad hoc networks without deployment constraints. Therefore they are not immediately tailored to describe the unique deployment features of vehicular networks.

Vehicular networks are expected to play a key role in improving traffic efficiency and safety in the near future~\cite{V2X5G, Karagiannis2011}. Using a planar two-dimensional \ac{PPP} to study their performance along orthogonal streets is not accurate in the high reliability regime~\cite{Haenggi2017}. An interference model for vehicular networks should naturally combine two spatial models; one for the road infrastructure and another for the locations of vehicles along each road. 

The Manhattan Poisson Line Process has been a popular model for the road network, where the resulting blocks might be filled in with buildings to resemble urban districts. In cellular systems, it has been shown that a user traveling on a street experiences discontinuous interference at the intersections~\cite{Baccelli2015}. In the absence of buildings, the interference from other roads can be easily mapped to interference from own road with a non-uniform density of users~\cite{Steinmetz2015, Alouini2016}. Recently, the Poisson Line Process has been used to model the random orientations of roads~\cite{Dhillon2017,Baccelli2017}. In an ad hoc setting, the intensities of roads and users have conflicting effects: Increasing the intensity of roads (while keeping fixed the intensity of vehicles per road) increases the interference while, increasing the intensity of vehicles  reduces the average link distance and improves coverage~\cite{Dhillon2017}. 

A common assumption in~\cite{Haenggi2017,Baccelli2015,Steinmetz2015, Alouini2016,Dhillon2017,Baccelli2017} is that the distribution of vehicles along a roadway follows the \ac{1D} \ac{PPP}. Similar assumption has been adopted  for  performance analysis over higher layers, e.g., the study in~\cite{Blaszczyszyn2013} jointly optimizes the transmission range and the transmission probability for maximizing transport capacity in linear networks with random access. There are some studies, e.g.~\cite{Alouini2016, Tong2016} using Mat\`{e}rn processes to approximate the density of simultaneous transmissions under the repulsive nature of IEEE 802.11p. The parent density is still \ac{PPP}. Finally, connectivity studies combining queueing theory with random geometric graphs often make a similar assumption for exponential distribution of  inter-arrivals~\cite{Chandrasekharamenon2012}. 

A great deal of transportation research since the early 1960's has recognized that the distribution of headway distance (the distance measured from the head of a vehicle to the head of its follower~\cite{Manual2000}, or simply the inter-vehicle distance) is not exponential under all circumstances. Different models were proposed to approximate the distribution of headway, with the accuracy of a particular model being dependent on the traffic status~\cite{Yin2009}. Empirical studies revealed  that the distribution of time headway (time difference between successive vehicles as they pass a point on the roadway~\cite{Manual2000}) is well-approximated by the log-normal \ac{PDF} under free flow~\cite{Daou1966, Greenberg1966} and by the log-logistic \ac{PDF} under congested flow~\cite{Yin2009}. Due to the mixed traffic conditions, Cowan has proposed not only single (exponential, shifted-exponential), but also mixed distribution models to describe the distribution of headways~\cite{Cowan1975}. 

To the best of our knowledge, apart from the exponential distribution, other headway models have not been incorporated into the interference analysis of vehicular networks. In~\cite{Yan2011}, the log-normal distribution along with the Fenton-Wilkinson method for approximating the distribution of multi-hop distances has been used to study the lifetime of a link. The randomness is due to the speed and headway, while fading and interference are neglected. 

Given a fixed and constant \ac{1D} intensity of users (or vehicles), the \ac{PPP} assumes that their locations are independent. Let us now consider a simple enhancement to the \ac{PPP}, which assumes that the headway distance is equal to the sum of a constant hardcore (or tracking) distance and an exponentially distributed \ac{RV}. The hardcore distance may model the average length of a vehicle plus a safety distance. Since the hardcore distance is assumed fixed and constant, the \ac{PDF} of headways becomes shifted-exponential. The motivation for this paper is to investigate how the first three moments of interference distribution behave under the shifted-exponential model. For instance, due to the fact that the deployment of interferers becomes more regular, the predicted interference is expected to have lower variance as compared to that originated from a \ac{PPP} of equal intensity.

The shifted-exponential distribution of headways,  makes the locations of vehicles correlated. The associated \ac{PCF} has been studied in the context of radial distribution function for hard spheres in statistical mechanics, see for instance~\cite{Salsburg1953, Sells1953}, and it has a complex form. As we will discuss later, the complexity of higher-order correlation functions does not allow us to calculate many more interference moments or bound the \ac{PGFL}~\cite{Baccelli2012}. Deriving the first moments of interference can serve as an intermediate step before approximating its \ac{PDF} with some simple function (with known Laplace transform) using, for instance, the method of moments. The contributions of this paper are:
\begin{itemize}
\item For small hardcore distance $c$ as compared to the mean inter-vehicle distance $\lambda^{-1}$, we show that the variance of interference at the origin can be approximated by the variance of interference due to a \ac{PPP} of equal intensity $\lambda$ scaled with $e^{-\lambda c}$. This model allows getting a quick insight on the impact of tracking distance $c$ on the variance under various traffic conditions. In addition, it shows that the distribution of interference becomes more concentrated around the mean in comparison with that due to a \ac{PPP} of intensity $\lambda$.  
\item We illustrate that large cell sizes $r_0$ and tracking distances $c$, modeling driving with high speeds at motorways, are associated with more concentrated distributions of interference (less coefficient of variation) and also more symmetric distributions (less skewness) in comparison with the distributions associated with urban microcells.
\item We study the variance of interference due to \ac{1D} infinite lattice to shed some light on the behavior of interference when the tracking distance becomes comparable to the mean inter-vehicle distance, approximating scenarios like flow of platoons of vehicles and traffic jams. We devise a simple, yet accurate, model approximating the variance under the assumptions of Rayleigh fading and small inter-point lattice distance as compared to the cell size.
\end{itemize}

The rest of the paper is organized as follows. In Section~\ref{sec:System}, we present the system set-up. In Section~\ref{sec:MomMeas}, we calculate the pair and higher-order correlations for the new deployment model. In Section~\ref{sec:MeanVar}, we calculate the mean, the variance and the skewness of interference. In Section~\ref{sec:VarApprox}, we derive closed-form approximations for the variance. In Section~\ref{sec:Lattice}, we study the extreme case where the interference originates from a lattice. In Section~\ref{sec:Conclusions}, we conclude the paper and discuss topics for future work. 

\section{System model}
\label{sec:System}
\begin{figure}[!t]
 \centering
  \includegraphics[width=3.5in]{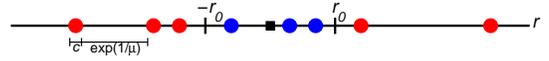}
 \caption{The vehicles are modeled as identical impenetrable disks. The vehicles outside of the cell (red disks) generate interference to the base station (black square). The rest (blue disks) are paired with the base station. In the figure, the tracking distance is illustrated equal to the diameter of the disk.}
 \label{fig:SystModel}
\end{figure}

Let us assume that the headway distance has two components: A constant tracking distance $c\!>\! 0$ and a free component following an exponential \ac{RV} with mean $\mu^{-1}$. This model degenerates to the time headway model M2 proposed by Cowan~\cite{Cowan1975}, if all vehicles move with the same constant speed towards the same direction. We study interference at a single snapshot. The base station is located at the origin, and the vehicles located in the interval $[-r_0,r_0]$ are associated to the base station, not contributing to interference level. The rest of the vehicles generate interference, see Fig.~\ref{fig:SystModel}.

While the performance evaluation of wireless cellular networks focuses mostly on the downlink coverage of a user, the uplink performance in emerging vehicular networks would be important too. Besides downlink transmissions for entertainment and infotainment services while on-board, uplink transmissions would be critical for traffic coordination, efficiency and safety. A valid study for the uplink should naturally incorporate power control, and the constraint that a single vehicle per antenna sector transmits at a time-frequency resource block, see~\cite{Andrews2013}. Due to the complex form of the \ac{PCF}, we leave this modeling details for future work. In this paper, we will get a preliminary insight into the impact of correlated user locations on the moments of interference in the uplink.

\begin{figure*}[!t]
 \centering
  \subfloat[$\lambda\!=\! 0.025 \text{m}^{-1}, c\!=\! 16 \text{m}$] {\includegraphics[width=2.5in]{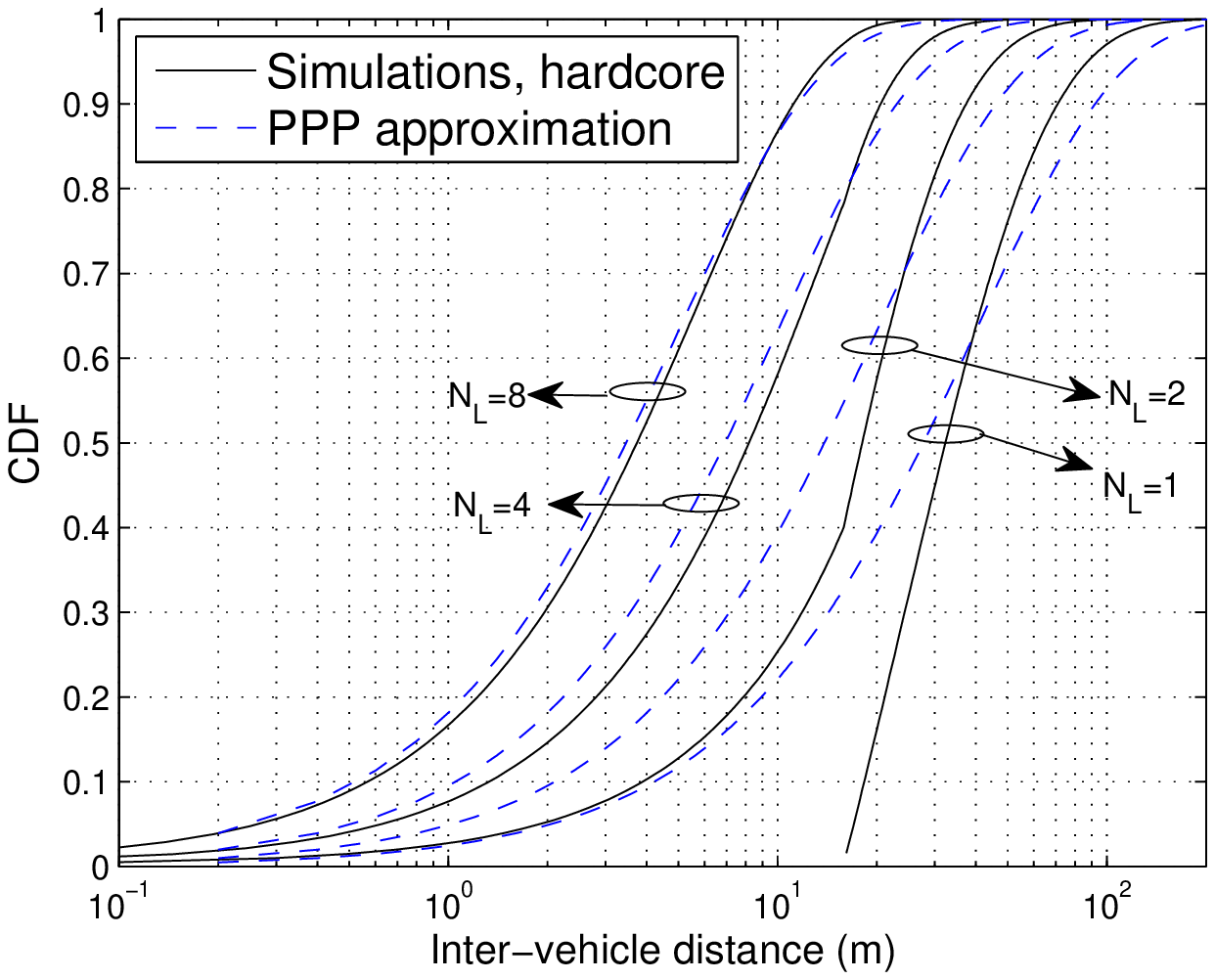}\label{fig:SuperpositionA}}\hfil \subfloat[$\lambda\!=\! 0.025 \text{m}^{-1}, c\!=\! 4 \text{m}$]{\includegraphics[width=2.5in]{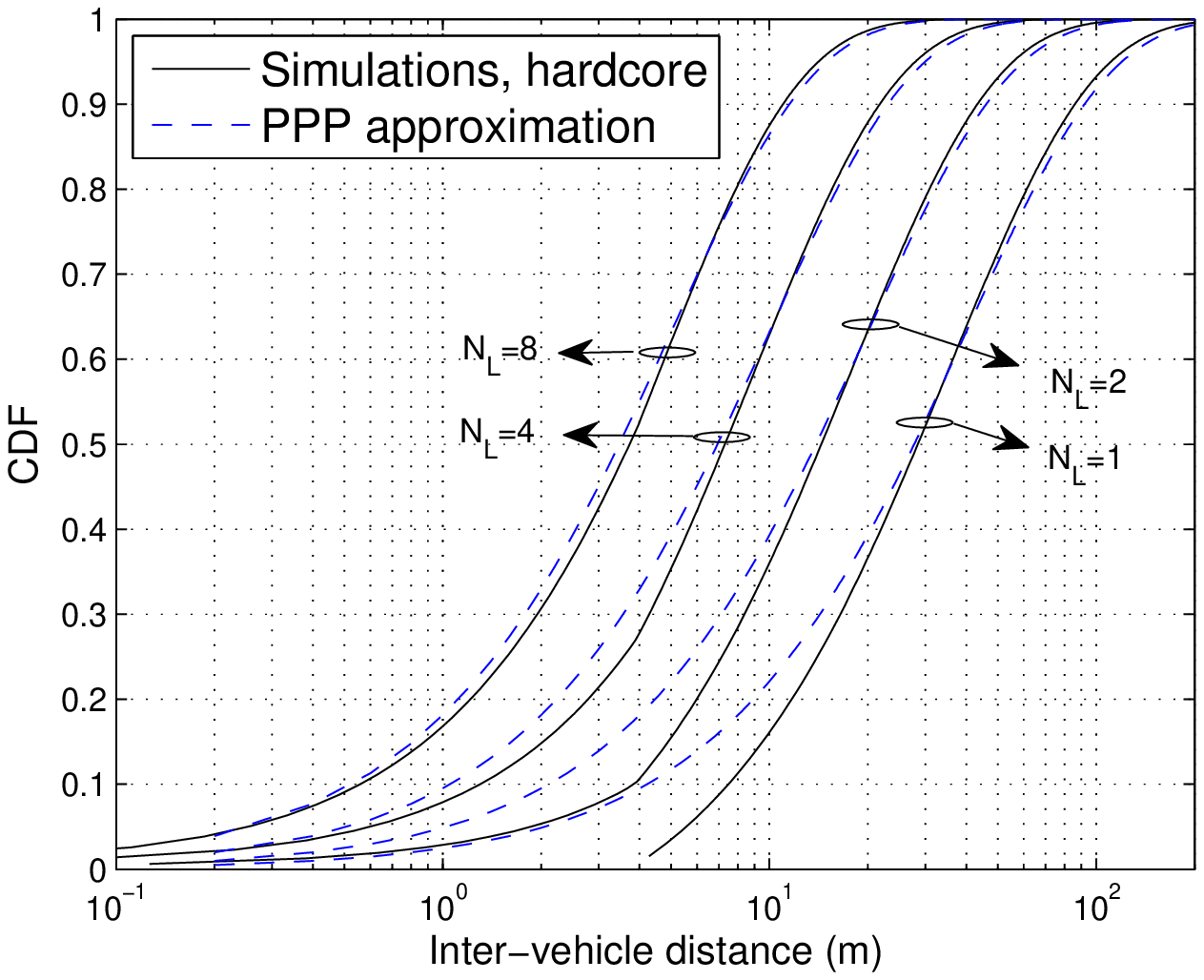}\label{fig:SuperpositionB}}
 \caption{Simulated \ac{CDF} of inter-vehicle distances resulting from the independent superposition (on the same line) of $N_L$ point processes of intensity $\lambda$ and hardcore distance $c$. The approximating \ac{CDF} using a \ac{PPP} of equivalent intensity $\lambda N_L$ is $F\!\left(x\right)\!=\!1-e^{-\lambda N_L x}$. $10\, 000$ simulation runs per curve.}
 \label{fig:Superposition}
\end{figure*}

Noting that the transmission range can be far greater than the width of a road, one may argue that in roads with multiple lanes, the distribution of inter-vehicle distances mapped onto a single line may still resemble an exponential. In Fig.~\ref{fig:SuperpositionA}, it is illustrated that for $\lambda c\!=\!0.4$ the distribution of inter-vehicle distances starts to converge to that due to a \ac{PPP} (of equivalent intensity) for more than eight lanes.  On the other hand, for smaller values of $\lambda c$, e.g., $\lambda c\!=\!0.1$ in Fig.~\ref{fig:SuperpositionB}, only four lanes might be enough to achieve quite good  convergence. Given that the product $\lambda c$ is fixed, the choice of parameters $\lambda,c$ does not impact the speed of convergence. We see that for two lanes, the sharp twist of the \ac{CDF} at inter-vehicle distance $x\!=\! c$ is still clear. Overall, the shifted-exponential model could be of use for roads with few lanes, e.g., bidirectional traffic streams with  restricted overtaking. In this kind of scenario, the model helps avoid unrealistically small headway distances predicted by the \ac{PPP} model with a high probability.

Regarding channel modeling, the propagation pathloss exponent is denoted by $\eta\!\geq\! 2$. The distance-based propagation pathloss function is $g\!\left(r\right)\!=\! \left|r\right|^{-\eta}$ for an interferer located at $r$ with $\left|r\right|\!>\!r_0$, and zero otherwise, to filter out vehicles inside the cell. The fast fading over each link is Rayleigh, and its multiplicative impact $h$ on the interference power is modeled by an exponential \ac{RV} with mean unity, $\mathbb{E}\left\{h\right\}\!=\!1$. The fading samples from different vehicles are \ac{i.i.d.}. The transmit power level is normalized to unity.  

\section{Moment measures}
\label{sec:MomMeas}
The simplest function incorporating the distance-dependent constraints of a point process is the second-order intensity measure   $\rho^{\left(2\right)}\!\left(x,y\right)$, or simply the \ac{PCF}. It describes the joint probability there are two points in the infinitesimal regions ${\rm d}x, {\rm d}y$ centered at $x$ and $y$ respectively. In order to express $\rho^{\left(2\right)}\!\left(x,y\right)$, we have to calculate the conditional probability there is a point at $y$ given a point at $x$.  For a \ac{PPP}, the locations of different points are independent, thus $\rho^{\left(2\right)}\!\left(x,y\right)\!=\!\lambda^2$, where $\lambda$ is the intensity. On the other hand, for the point process considered here the distance distribution between neighbors is shifted-exponential with positive shift $c$. Next, we show how to calculate $\rho^{\left(2\right)}\!\left(x,y\right)$ for this deployment rule. We will also generalize the calculation for the $n$-th order correlation function defined over $n-$tuples of points; needed in the calculation of the $n$-th moment of interference.

Due to the stationarity of the point process, the \ac{PCF} depends on the distance separation between $x$ and $y$. Let us assume $y\!>\!x\!>\! 0$ and denote by $\rho_k^{\left(2\right)}\!\!\left(y,x\right), k\!\in\!\mathbb{N}$, the branch of the \ac{PCF} for  $y\!\in\!\left(x\!+\!kc,x\!+\!\left(k\!+\!1\right)c\right)$. Since two vehicles are separated at least by the tracking distance, the \ac{PCF} becomes zero for distances smaller than $c$, and thus  $\rho_0^{\left(2\right)}\!\!\left(y,x\right)\!=\! 0$. For distance separation between $c$ and $2c$, no other vehicles can be located in-between. Therefore $\rho_1^{\left(2\right)}\!\!\left(y,x\right)\!=\!\lambda\mu e^{-\mu\left(y-x-c\right)}$, where $\lambda\mu {\rm d}x{\rm d}y$ is the probability that two vehicles are located in the infinitesimal regions ${\rm d}x, {\rm d}y$, and  $e^{-\mu\left(y-x-c\right)}$ is the probability that no other vehicle is located in $\left(x\!+\!c,y\right)$. For distance separation between $2c$ and $3c$, at most one vehicle can be located in-between, and the \ac{PCF} consists of two terms.
\[
\begin{array}{ccl}
 \displaystyle \rho_2^{\left(2\right)}\!\!\left(y,x\right) \!\!\!\!\!&=&\!\!\!\!\! \displaystyle \frac{\lambda\mu}{e^{\mu\left(y-x-c\right)}}  +   \lambda\!\int_{x+c}^{y-c}\!\!\!\!\!\!\! \mu e^{-\mu\left(z-x-c\right)} \mu  e^{-\mu\left(y-z-c\right)} {\rm d}z \\ 
  \!\!\!\!\!&=&\!\!\!\!\! \displaystyle \frac{\lambda\mu}{e^{\mu\left(y-x-c\right)}}  + \frac{\lambda\mu^2 \left(y\!-\!x\!-\!2c\right)}{e^{\mu\left(y-x-2c\right)}},
\end{array}
\]
where $\mu e^{-\mu\left(z-x-c\right)} {\rm d}z$ is the probability that a vehicle is located in the region ${\rm d}z$ centered at $z\!\in\!\left(x\!+\!c,y\!-\!c\right)$.
\begin{figure}[!t]
 \centering
  \includegraphics[width=3.0in]{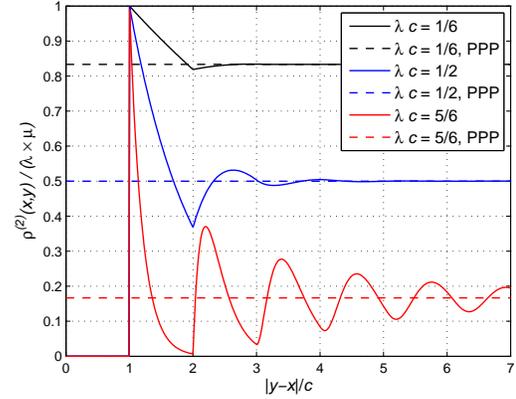}
 \caption{Normalized \ac{PCF} $\rho^{\left(2\right)}\!\left(x,y\right)/\left(\lambda \mu\right)$ with respect to the normalized distance $|y\!-\!x|/c$.  The dashed lines correspond to $\rho^{\left(2\right)}\!\left(x,y\right)\!=\!\lambda^2$, or equivalently, $\rho^{\left(2\right)}\!\left(x,y\right)/\left(\lambda \mu\right)\!=\!1\!-\!\lambda c$.}
 \label{fig:CorrFunc}
\end{figure}

Following the same reasoning, when the distance $\left(y\!-\!x\right)$ is between $3c$ and $4c$, there are at most two vehicles in-between, and the \ac{PCF} has three terms. The way to calculate the probabilities for zero and one vehicle between $x$ and $y$ has been shown. It remains to calculate the probability there are two vehicles. Let us assume that the vehicles are located at $z_1\!<\! z_2$. Then $z_1\!\in\!\left(x\!+\!c,y\!-\!2c\right)$ and $z_2\!\in\!\left(z_1\!+\!c,y\!-\!c\right)$. The  probability that four vehicles are located at $x\!<\!z_1\!<\!z_2\!<\!y$ is  
\[ 
\lambda\!\int_{x+c}^{y-2c}\!\!\!\int_{z_1+c}^{y-c} \!\!\!\!\! \mu^3 e^{-\mu\left(z_1-x-c\right)} e^{-\mu\left( z_2-z_1-c\right)} e^{-\mu\left( y-z_2-c\right)}\!{\rm d}z_2 {\rm d}z_1.
\]

\noindent 
After carrying out the integration and summing up,  
\[
\begin{array}{ccl}
 \displaystyle \rho_3^{\left(2\right)}\!\!\left(y,x\right) \!\!\!\!\!&=&\!\!\!\!\!  
 \displaystyle \frac{\lambda\mu}{e^{\mu\left(y-x-c\right)}} \!+\! \frac{\lambda\mu^2 \! \left(y\!-\!x\!-\!2c\right)}{e^{\mu\left(y-x-2c\right)}}  \!+\! \frac{\lambda\mu^3 \! \left(y\!-\!x\!-\!3c\right)^2}{2 e^{\mu\left(y-x-3c\right)}}. 
\end{array}
\]

\noindent 
Similarly, we can compute $\rho_k^{\left(2\right)}\!\!\left(y,x\right)$ for larger $k$. 
\begin{equation}
\label{eq:rho}
\rho_k^{\left(2\right)}\!\!\left(y,x\right) \!\!=\!\! \Bigg\{ \!\!\!\! \begin{array}{ccl}\lambda \!\! \sum\limits_{j=1}^k \!\! \frac{\mu^j \left(y-x-jc\right)^{j-1}}{\Gamma\!\left(j\right) e^{\mu\left(y-x-jc\right)}}, \!\!\!\!\!\!& &\!\!\!\!\!\! y\!\in\!\left(x\!\!+\!kc, x\!\!+\!\left(k\!+\!1\right)\!c\right) \\ 0, \!\!\!\!\!\!& &\!\!\!\!\!\! \text{otherwise}, \end{array} 
\end{equation}
where $k\!\geq\! 1$ and $\Gamma\!\left(j\right)\!=\!\left(j\!-\!1\right)!$ 

Obviously, $\rho^{\left(2\right)}\!\left(x,y\right)\!=\!\sum_{k=0}^\infty \!\rho_k^{\left(2\right)}\!\!\left(y,x\right), y>x$. For $y\!<\!x$, we just need to interchange $x$ and $y$ in~\eqref{eq:rho}. This \ac{PCF} has also been studied in the context of statistical mechanics to describe the density variations of particles for \ac{1D} hardcore fluids as compared to the ideal (\ac{PPP}) fluid~\cite{Salsburg1953, Sells1953}. The derivation of~\eqref{eq:rho} in~\cite{Salsburg1953, Sells1953} is carried out using thermodynamic equations of state. It is the probability of finding a particle at a distance $\left(y\!-\!x\right)$ from an arbitrary fixed particle at $x$. We have used basic probability theory instead, to highlight the constraints introduced by the deployment rule. The Laplace tranform of~\eqref{eq:rho} is available in~\cite[pp.~5]{Mattis1993}. In Fig.~\ref{fig:CorrFunc}, we depict the normalized \ac{PCF}. For small $\lambda c$, the function decorrelates within few multiples of $c$. For increasing $\lambda c$, the locations of two vehicles can remain correlated over larger ranges. 

Let us consider $n$ points on the real line, $x_1,x_2,\ldots x_n$, in increasing order. According to~\cite[Eq. (27)]{Salsburg1953}, the higher-order intensity measure $\rho^{\left(n\right)},\, n\!\geq\!3$ for the shifted-exponential deployment has the following form, $\rho^{\left(n\right)}\!\left(x_1,x_2,\ldots x_n \right)\!=\! \frac{1}{\lambda^{n-2}}\prod_{i=1}^{n-1}\rho^{\left(2\right)}\!\left(x_{i+1}\!-\!x_i \right)$. For instance, the third-order intensity that describes the probability to find a triple of distinct vehicles at $x,y$ and $w$, is  
\[
\rho^{\left(3\right)}\!\left(x,y,w\right) = \frac{1}{\lambda} \, \rho^{\left(2\right)}\!\left(x,y\right) \rho^{\left(2\right)}\!\left(y,w\right).
\]

The performance assessment of wireless networks commonly utilizes the coverage probability as a metric, i.e., the probability (over the ensemble of all  network states) that the Signal-to-Interference-and-Noise ratio is larger than a threshold. Even if the distribution of interference is unknown, the coverage probability could be computed, provided that the \ac{PGFL} of the point process generating the interference is available. For a non-Poissonian point process, this is in general difficult to calculate. In addition, we saw that the $n-$th order intensity $\rho^{\left(n\right)}$ has increasing complexity for increasing $n$. Because of that, we could not simplify the multi-dimensional integrations in~\cite[Eq. (14)]{Baccelli2012} to obtain tight bounds for the coverage probability. In order to bypass the calculation of the \ac{PGFL}, we may approximate the interference by some well-known \ac{PDF} with parameters, selected for instance using the method of moments. In that case, even two or three moments of interference might be sufficient for a good fit.  Some discussion about the \ac{PDF} of aggregate interference with a guard zone around the receiver can be found in~\cite[Section III]{Haenggi2013}.  In this regard, we show next how to  calculate the first three moments of interference. 

\section{Moments of interference}
\label{sec:MeanVar}
\begin{figure}[!t]
 \centering
  \includegraphics[width=3.0in]{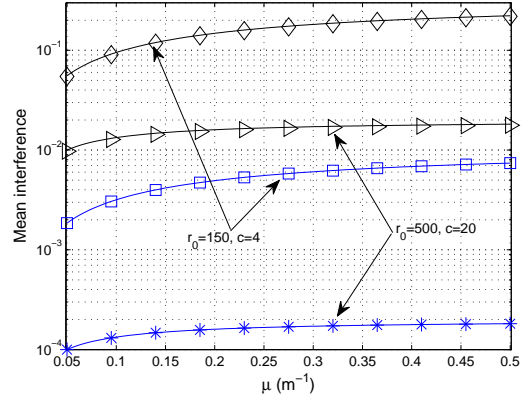}
 \caption{Mean interference with respect to the random part $\mu$ of the intensity of vehicles. $50\, 000$ simulation runs per marker over a line segment of $40$ km. The solid lines correspond to~\eqref{eq:EI} validated against the simulations (markers). Given the deployment scenario, the higher mean interference corresponds to pathloss exponent $\eta\!=\!2$, and the lower to $\eta\!=\!3$.}
 \label{fig:Mean}
\end{figure}
The mean interference at the origin can be calculated using the Campbell's Theorem for stationary processes~\cite{Mecke2013}. Given the traffic parameters $\mu,c$, the intensity $\lambda$ of vehicles is constant and equal to  $\lambda^{\!-1}\!=\! c + \mu^{\!-1}$, or $\lambda\!=\!\frac{\mu}{1+c\mu}$~\cite{Cowan1975}. After averaging the distance-based propagation pathloss over the intensity of interferers, we get the mean interference level. 
\begin{equation}
\label{eq:EI}
\mathbb{E}\!\left\{\mathcal{I}\right\} = 2 \lambda  \mathbb{E}\!\left\{h\right\} \!\int_{r_0}^{\infty}\!\!\!  g\!\left(r\right) {\rm d}r = \frac{2\lambda r_0^{1-\eta}}{\eta-1}, 
\end{equation}
where the factor two is due to vehicles at negative half-axis. 

In Fig.~\ref{fig:Mean} we consider two scenarios: (i)  tracking distance $c\!=\!4$ m with cell size $r_0\!=\!150$ m modeling vehicular networks in urban street microcells, and (ii) $c\!=\!20$ m and $r_0\!=\!500$ m modeling driving at higher speeds (hence the larger tracking distance) in motorway macrocells. We illustrate the mean interference for increasing traffic conditions, i.e., increasing the random part $\mu$ of the deployment model. For each scenario, we depict the interference level for two channel models, $\eta\!=\!2$ and $\eta\!=\!3$. The large cell size in conjunction with the large tracking distance makes the mean interference level less sensitive to the random part $\mu$ of the traffic intensity.

The mean interference due to a \ac{PPP} of intensity $\lambda$ is still given by~\eqref{eq:EI}. However, this is not the case for higher moments. We shall see that different hardcore distances $c$ result in different variance and skewness of interference while keeping the intensity $\lambda$ of vehicles fixed by varying $\mu\!=\!\frac{\lambda}{1\!-\!\lambda c}$. 

The second moment of interference accepts contributions not only from a single vehicle but also from pairs. 
\begin{equation}
\label{eq:EI2}
\mathbb{E}\!\left\{\mathcal{I}^2\right\}  =  2 \lambda \! \int \!\!g^2\!\left(r\right) {\rm d}r + \int \!\! g\!\left(x\right)g\!\left(y\right)\rho^{\left(2\right)}\!\left(x,y\right) {\rm d}x {\rm d}y, 
\end{equation}
where the factor two in front of the first term comes from the second moment of a unit-mean exponential \ac{RV},  $\mathbb{E}\left\{h^2\right\}\!=\!2$. 

In order to calculate $S\!=\! \int\!  g\!\left(x\right)\! g\!\left(y\right)\!\rho^{\left(2\right)}\!\left(x,y\right) {\rm d}x {\rm d}y$, we substitute equation~\eqref{eq:rho} into it, remembering to interchange $x$ and $y$ for $x\!>\!y$. 
\begin{equation}
\label{eq:Ir0}
\begin{array}{ccl}
S \!\!\!\!\!&=&\!\!\!\!\! \displaystyle 2\sum\limits_{k=1}^\infty\!\!\int_{r_0}^\infty\!\!\!\!\int_{x+kc}^{x+\left(k+1\right)c} \!\!\!\!\!\!\!\!\!\!\!\!\!\!\!\!\!\!\!\! g\!\left(x\right) g\!\left(y\right) \rho_k^{(2)}\!\!\left(y,x\right) {\rm d}y {\rm d}x \, + \\ & & \displaystyle 2\sum\limits_{k=1}^\infty\!\!\int_{r_0}^\infty\!\!\!\!\int_{x-\left(k+1\right)c}^{x-kc} \!\!\!\!\!\!\!\!\!\!\!\!\!\!\!\!\!\!\!\! g\!\left(x\right) g\!\left(y\right) \rho_k^{(2)}\!\!\left(x,y\right) {\rm d}y {\rm d}x \\ 
 \!\!\!\!\!&=&\!\!\!\!\! \displaystyle 2\lambda\!\sum\limits_{k=1}^\infty \!\! \int_{r_0}^\infty\!\!\!\!\int_{x+kc}^{x+\left(k+1\right)c}  \!\!\!\!\!\!\!\!\!\! \!\!\!\!\!\!\!\!\!\!\!\!\! g\!\left(x\right)\!g\!\left(y\right) \!\!\sum\limits_{j=1}^k\!\frac{\mu^j \! \left(y\!-\!x\!-\!jc\right)^{j-1}}{\Gamma\!\left(j\right) e^{\mu\left( y-x-jc\right)}}  {\rm d}y{\rm d}x + \\ 
{} \!\!\!\!\!& &\!\!\!\!\!  \displaystyle 2\lambda\!\sum\limits_{k=1}^\infty \!\! \int_{r_0}^\infty\!\!\!\!\int_{x-\left(k+1\right)c}^{x-kc}  \!\!\!\!\!\!\! \!\!\!\!\!\!\!\!\!\!\!\!\! g\!\left(x\right)\!g\!\left(y\right) \!\! \sum\limits_{j=1}^k\!\frac{\mu^j \! \left(x\!-\!y\!-\!jc\right)^{j-1}}{\Gamma\!\left(j\right) e^{\mu\left( x-y-jc\right)}}   {\rm d}y {\rm d}x, 
\end{array}
\end{equation}
where the factor two is added to account for $x\!\leq\!-r_0$.

The calculation of $S$ involves double integration, infinite sums and requires to filter out the vehicles within the cell. In order to simplify it, we note that for increasing distance separation, the \ac{PCF} becomes progressively  equal to $\lambda^2$. Let us assume an integer $m\!\geq\!2$, and approximate $\rho_k^{(2)}\!\left(y,x\right)\!\approx\!\lambda^2,\, \forall k\!\geq\!m$ (similar for $x\!>\!y$). From the first equality in~\eqref{eq:Ir0} we have    
\[
\begin{array}{ccl}
S \!\!\!\!\!&\approx&\!\!\!\!\! \displaystyle 2\sum\limits_{k=1}^{m-1}\!\!\int_{r_0}^\infty\!\!\!\!\int_{x+kc}^{x+\left(k+1\right)c} \!\!\!\!\!\!\!\!\!\!\!\!\!\!\!\!\!\!\!\! g\!\left(x\right) g\!\left(y\right) \rho_k^{(2)}\!\!\left(y,x\right) {\rm d}y {\rm d}x \,\,\, + \\ & & \displaystyle 2\sum\limits_{k=1}^{m-1}\!\!\int_{r_0}^\infty\!\!\!\!\int_{x-\left(k+1\right)c}^{x-kc} \!\!\!\!\!\!\!\!\!\!\!\!\!\!\!\!\!\!\!\! g\!\left(x\right) g\!\left(y\right) \rho_k^{(2)}\!\!\left(x,y\right) {\rm d}y {\rm d}x \,\,\, + \\ 
\!\!\!\!\!& &\!\!\!\!\! \displaystyle 2\lambda^2\!\! \sum\limits_{k=m}^\infty\!\! \left(  \int_{r_0}^\infty\!\!\!\!\int_{x+kc}^{x+\left(k+1\right)c} \!\!\!\!\!\!\!\!\!\!\!\!\!\!\!\!\!\!\!\! g\!\left(x\right) g\!\left(y\right)  {\rm d}y {\rm d}x \!+\!\! \int_{r_0}^\infty\!\!\!\!\int_{x-\left(k+1\right)c}^{x-kc} \!\!\!\!\!\!\!\!\!\!\!\!\!\!\!\!\!\!\!\! g\!\left(x\right) g\!\left(y\right)  {\rm d}y {\rm d}x \!\! \right)\!.
\end{array}
\]

\noindent 
The last line above can also be written as 
\[
2\lambda^2\!\left( \int_{r_0}^\infty\!\!\!\!\int_{x+mc}^\infty \!\!\!\!\!\!\!\!\!\!\! g\!\left(x\right) g\!\left(y\right)  {\rm d}y {\rm d}x \!+\!\! \int_{r_0}^\infty\!\!\!\!\int_{-\infty}^{x-mc} \!\!\!\!\!\!\!\!\!\!\!\!\! g\!\left(x\right) g\!\left(y\right)  {\rm d}y {\rm d}x\right)\!.
\]

Using the exact expression of the \ac{PCF} up to $\left(m\!+\!1\right)$ comes at the cost of calculating the integrals $\int_{r_0}^\infty\!\!\int_{x+mc}^{x+\left(m+1\right)c} \!\!\!\! g\!\left(x\right) g\!\left(y\right) \rho_m^{(2)}\!\left(y,x\right) {\rm d}y {\rm d}x$. Therefore the higher the $m$ is, the higher is the penalty for improving the accuracy. 

\noindent 
For $m\!=\!2$ we get
\begin{equation}
\begin{array}{ccl}
\label{eq:Iapprox}
S \!\!\!\!\!&\approx&\!\!\!\!\! \displaystyle 2\lambda\mu \!\! \int_{r_0}^\infty\!\!\! \bigg( \! \int\limits_{x+c}^{x+2c}\!\!\!\!\! \frac{g\!\left(x\right)\! g\!\left(y\right)}{e^{\mu\left( y-x-c\right)}}{\rm d}y  \!+\!\!  \!\!\!\int\limits_{x-2c}^{x-c}\!\!\!\!\! \frac{g\!\left(x\right)\! g\!\left(y\right)}{e^{\mu\left( x-y-c\right)}}{\rm d}y\bigg) {\rm d}x \, + \\ 
{} \!\!\!& &\!\!\! \displaystyle  2\lambda^2 \!\!\! \int_{r_0}^\infty \!\!\! \bigg( \int_{x+2c}^\infty\!\!\!\!\!\!\!\!\! g\!\left(x\right)\! g\!\left(y\right) {\rm d}y \!+\!\! \int_{-\infty}^{x-2c}\!\!\!\!\!\!\!\!\! g\!\left(x\right)\! g\!\left(y\right) {\rm d}y \bigg) {\rm d}x. 
\end{array}
\end{equation}
\begin{figure*}[!t]
 \centering
  \subfloat[Coefficient of variation] {\includegraphics[width=2.5in]{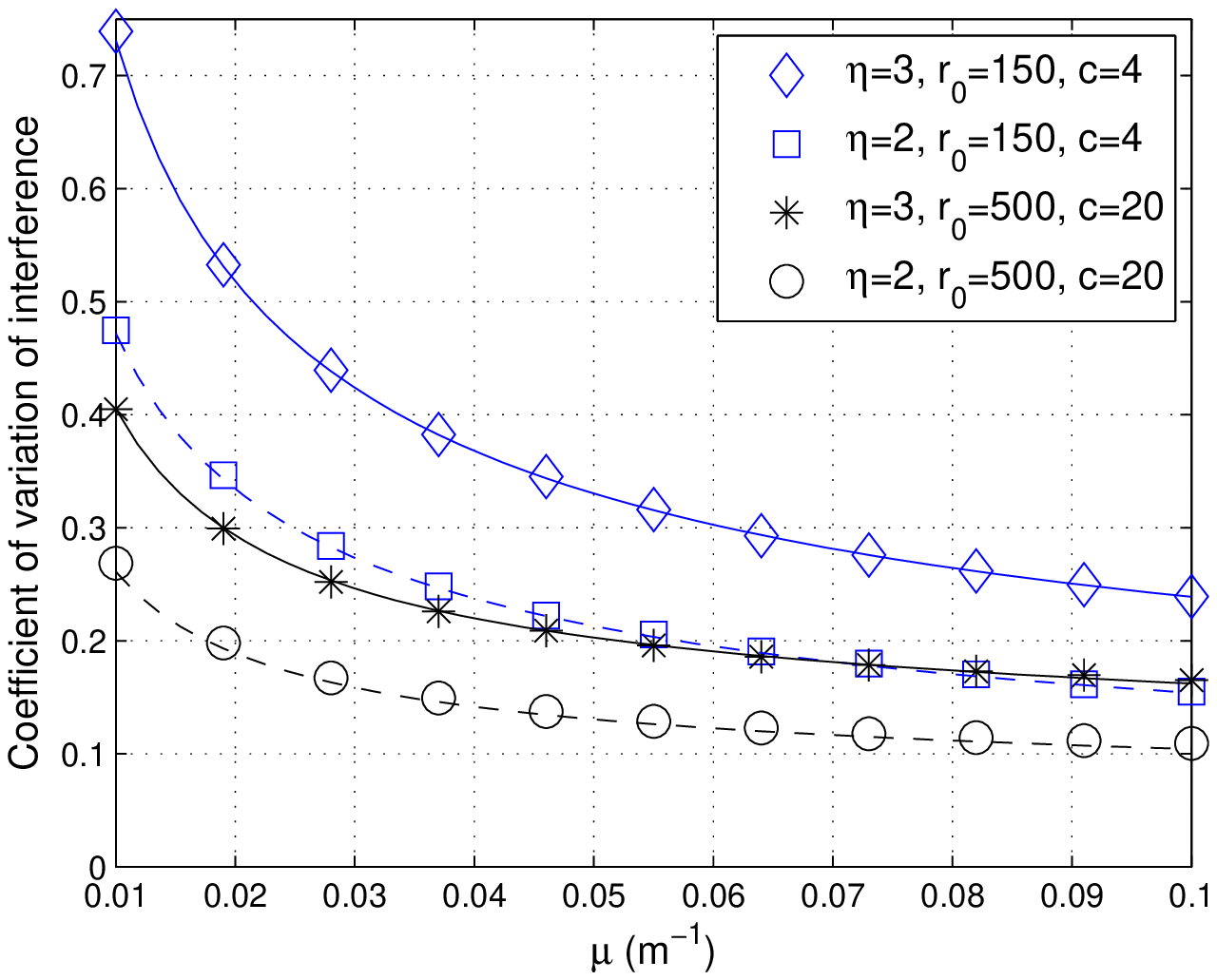}\label{fig:StdMu}}\hfil \subfloat[Skewness]{\includegraphics[width=2.5in]{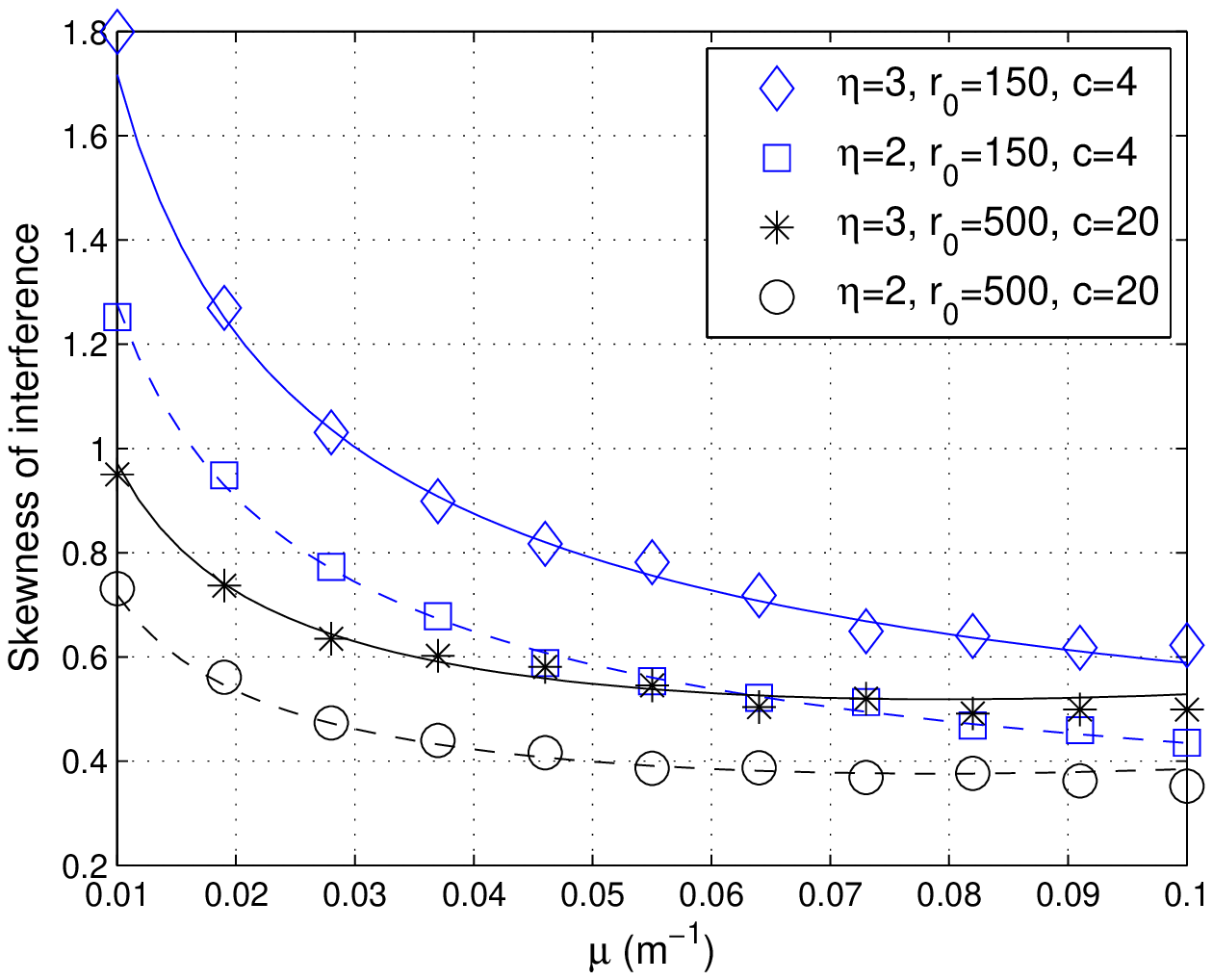}\label{fig:SkewnessMu}}
 \caption{Statistics of interference with respect to the random part $\mu$ of the deployment. The approximations (solid and dashed lines) use $m\!=\!2$ and numerical integration for the  terms $S, S', S_1'', S_2''$. The approximations are validated against the simulations (markers). $10^5$ simulation runs per marker. The simulations are carried out over a line segment of $40$ km.}
 \label{fig:StdSkewMu}
\end{figure*}

In order to see the complications in the calculation of higher interference moments, we show the calculation of the third moment, which accepts contributions from a single user, from pairs and from triples of users. 
\begin{equation}
\label{eq:EI3}
\begin{array}{ccl}
\mathbb{E}\!\left\{\mathcal{I}^3\right\} \!\!\!\!\!\!\!&=&\!\!\!\!\! \displaystyle  6\lambda\!\! \int \!\! g^3\!\!\left(r\right){\rm d}r \!+\! 6\!\!\int\!\! g^2\!\!\left(x\right)g\!\left(y\right)\rho^{(2)}\!\!\left(x,y\right)\! {\rm d}x{\rm d}y  + \\ \!\!\!\!\!& & \displaystyle \int\!\! g\!\left(x\right)g\!\left(y\right)g\!\left(w\right) \rho^{(3)}\!\left(x,y,w\right) {\rm d}x{\rm d}y{\rm d}w, 
\end{array}
\end{equation}
where the factor six in the first term comes from the third moment of an exponential \ac{RV}, $\mathbb{E}\!\left\{h^3\right\}\!=\!6$, and the same factor in the second term comes from multiplying the second moment of an exponential \ac{RV}, $\mathbb{E}\!\left\{h^2\right\}\!=\!2$, with the three possible ways to select a pair out of a triple of users.

We still approximate the \ac{PCF} by $\lambda^2$ beyond $2c$. The term  $S'\!=\!\int\! g^2\!\!\left(x\right)g\!\left(y\right)\rho^{(2)}\!\!\left(x,y\right) {\rm d}x{\rm d}y$ can be expressed similar to the term $S$ in~\eqref{eq:Iapprox}.
\[
\begin{array}{ccl}
S' \!\!\!\!\!&\approx&\!\!\!\!\! \displaystyle 2\lambda\mu \!\! \int_{r_0}^\infty\!\!\! \bigg( \! \int\limits_{x+c}^{x+2c}\!\!\!\!\! \frac{g^2\!\!\left(x\right)\! g\!\left(y\right)}{e^{\mu\left( y-x-c\right)}}{\rm d}y  \!+\!\!  \!\!\!\int\limits_{x-2c}^{x-c}\!\!\!\!\! \frac{g^2\!\left(x\right)\! g\!\left(y\right)}{e^{\mu\left( x-y-c\right)}}{\rm d}y\bigg) {\rm d}x \, + \\ 
{} \!\!\!& &\!\!\! \displaystyle  2\lambda^2 \!\!\! \int_{r_0}^\infty \!\!\! \bigg( \int_{x+2c}^\infty\!\!\!\!\!\!\!\!\! g^2\!\!\left(x\right)\! g\!\left(y\right) {\rm d}y \!+\!\! \int_{-\infty}^{x-2c}\!\!\!\!\!\!\!\!\! g^2\!\!\left(x\right)\! g\!\left(y\right) {\rm d}y \bigg) {\rm d}x. 
\end{array}
\]

Calculating $S''\!=\!\int\!\! g\!\left(x\right)g\!\left(y\right)g\!\left(w\right) \rho^{(3)}\!\left(x,y,w\right) {\rm d}x{\rm d}y{\rm d}w$ is more tedious because the third-order correlation is equal to the product of \acp{PCF}, $\rho^{\left(3\right)}\!\left(x,y,w\right) = \frac{1}{\lambda} \rho^{\left(2\right)}\!\left(x,y\right) \rho^{\left(2\right)}\!\left(y,w\right)$. Fortunately, the pathloss function $g\!\left(\cdot\right)$ is common for the three users. Therefore it suffices to calculate $S''$ for a particular order and scale the result by six. 
\[
\begin{array}{ccl}
S'' \!\!\!\!\!&\approx&\!\!\!\!\! \displaystyle \frac{6}{\lambda} \int\!\! g\!\left(x\right)g\!\left(y\right)g\!\left(w\right) \rho^{(2)}\!\left(x,y\right) \rho^{(2)}\!\left(y,w\right) {\rm d}x{\rm d}y{\rm d}w + \\ \!\!\!\!\!& & \displaystyle 6\lambda \int\!g\!\left(x\right){\rm d}x \int g\!\left(y\right)g\!\left(w\right) \rho^{(2)}\!\left(y,w\right) {\rm d}y{\rm d}w, 
\end{array}
\]
where the first term corresponds to ordered users $x\!<\!y\!<\!w$ at the same side of the cell, and the second term describes the case with the user $x$ (approximately) uncorrelated to the locations of $y,w$ ($y\!<\!w$) because it is placed at the opposite side, thus $\rho^{(2)}\!\left(x,y\right)\!\approx\!\lambda^2$.  

Since we consider the exact expression for the \ac{PCF} up to $2c$, the first term of $S''$ above, let us denote it by $S_1''$, can be separated into four terms describing the distance separations (closer or further than $2c$) between the users of each pair $\left\{y,w\right\}$ and $\left\{x,y\right\}$. 
\[
\begin{array}{ccl}
S_1''\!\!\!\!\!\!&\approx&\!\!\!\!\!\! \displaystyle 12\lambda^2\!\!\!\int\limits_{r_0}^\infty\!\int\limits_{x+2c}^\infty\!\!\!\!\left(\int\limits_{y+c}^{y+2c}\!\!\!\!\! \frac{\mu g\!\left(\!w\!\right) {\rm d}w}{e^{\mu\left(w-y-c\right)}}  \!+\!\!\!\! \int\limits_{y+2c}^\infty \!\!\!\!\!\lambda g\!\left(\!w\!\right){\rm d}w\!\!\right)\!\!  g\!\left(\!x\!\right) \!g\!\left(\!y\!\right) \! {\rm d}y {\rm d}x \, + \\ 
\!\!\!\!\!\!& &\!\!\!\!\!\!\!\!\! \displaystyle 12\lambda\mu\!\!\!\int\limits_{r_0}^\infty\!\int\limits_{x+c}^{x+2c}\!\!\!\left(\int\limits_{y+c}^{y+2c}\!\!\!\!\! \frac{\mu g\!\left(\!w\!\right) {\rm d}w}{e^{\mu\left(w-y-c\right)}}  \!+\!\!\!\! \int\limits_{y+2c}^\infty \!\!\!\!\! \lambda g\!\left(\!w\!\right){\rm d}w\!\!\right)\!\!\!\frac{g\!\left(\!x\!\right) g\!\left(\!y\!\right)}{e^{\mu\left(y-x-c\right)}} \, {\rm d}y {\rm d}x, 
\end{array}
\]
where the factor two is due to symmetry, i.e., the three users are located at the negative half-axis.
\begin{figure*}[!t]
 \centering
  \subfloat[Standard deviation] {\includegraphics[width=2.5in]{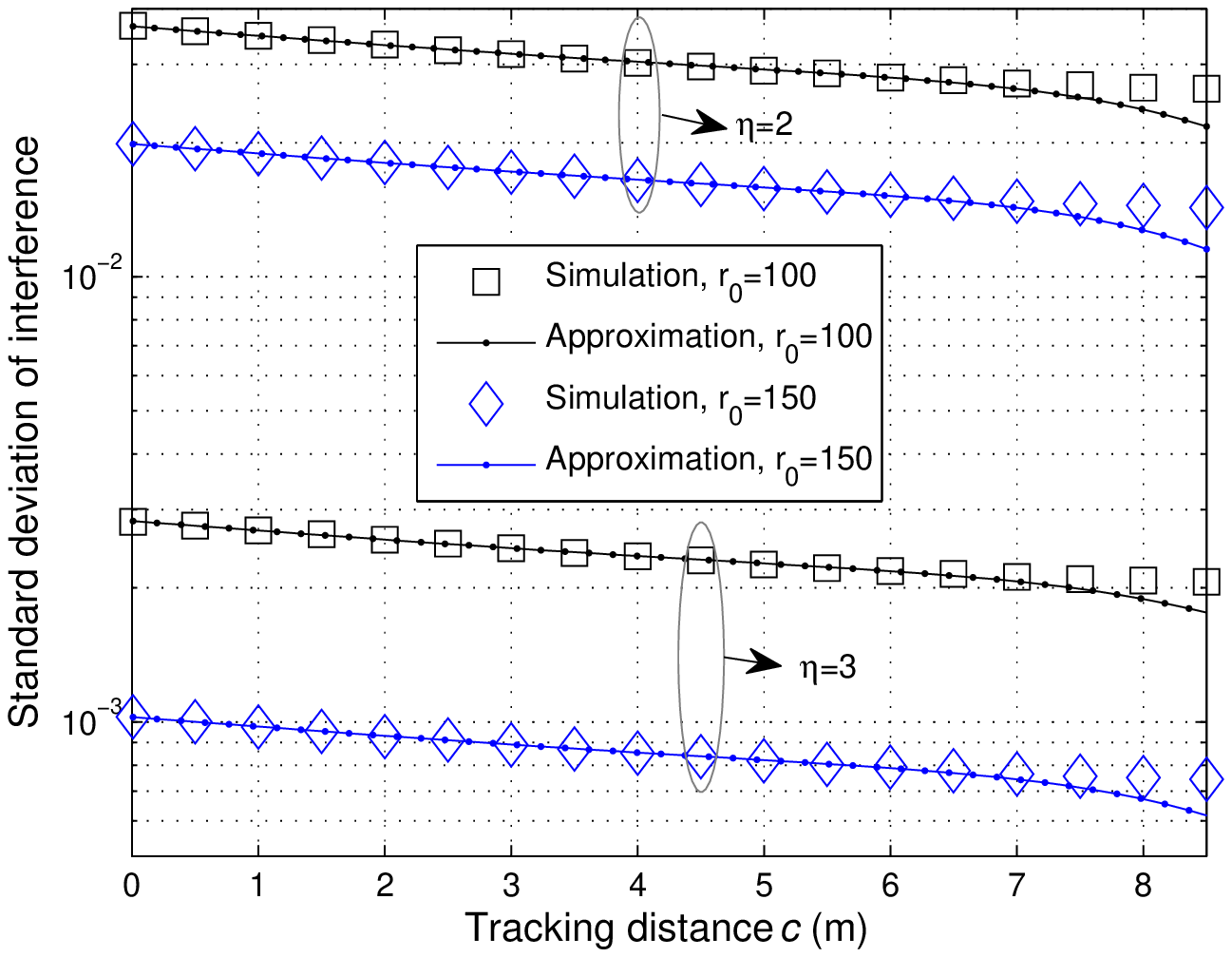}\label{fig:StdSkewa}}\hfil \subfloat[Skewness]{\includegraphics[width=2.5in]{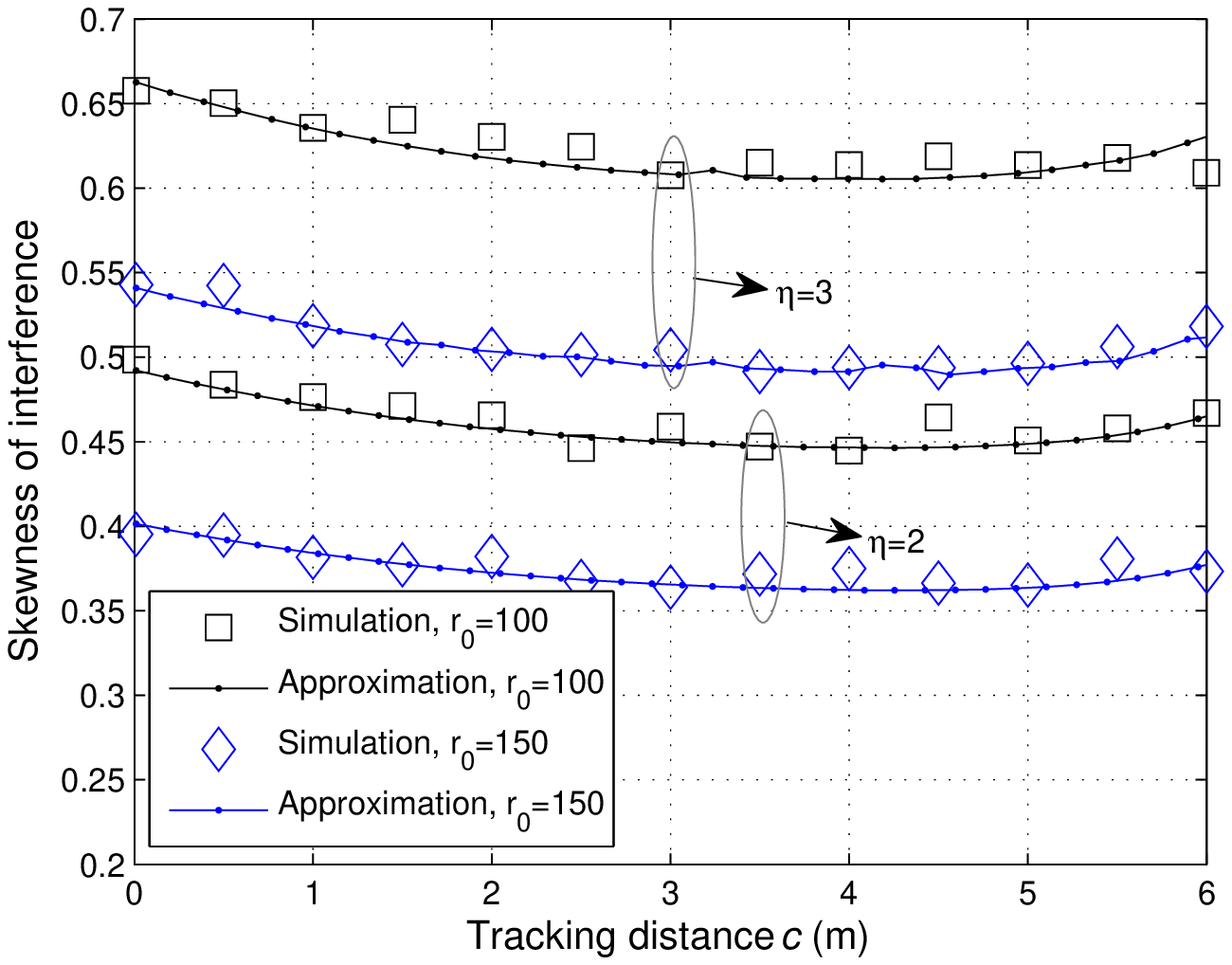}\label{fig:StdSkewb}}
 \caption{Statistics of interference for a fixed intensity of vehicles $\lambda\!=\!0.1 {\text{m}}^{-1}$. The calculations using integral-based approximation for $S, S', S_1'', S_2''$  for $m\!=\!2$ are validated against the simulations (markers). $10^5$ simulation runs per marker. The simulations are carried out over a line segment of $40$ km.}
 \label{fig:StdSkew}
\end{figure*}

We continue with the second term of $S''$, let us denote it by $S_2''$. In the expression of $S_2''$ the users $y$ and $w$ are already ordered and placed at the same side of the cell. After using the approximation for the \ac{PCF} beyond $2c$ we get 
\[
\begin{array}{ccl}
S_2''\!\!\!&\approx&\!\!\! \displaystyle 12\lambda^2\!\!\!\int\nolimits_{r_0}^\infty \!\! g\!\left(x\right) \! {\rm d}x \Bigg( \lambda\!\! \int\nolimits_{r_0}^\infty\!\!\!\!\int_{y+2c}^\infty \!\!\!\! g\!\left(y\right) \! g\!\left(w\right) \!{\rm d}w {\rm d}y \, + \\ & & \displaystyle \mu\! \int\nolimits_{r_0}^\infty\!\!\!\!\int_{y+c}^{y+2c} \!\! \frac{g\!\left(y\right) \! g\!\left(w\right) {\rm d}w {\rm d}y}{e^{\mu\left(w-y-c\right)}} \Bigg), 
\end{array}
\]
where the factor two is again due to symmetry, i.e., the sides (with respect to the cell) of the user $x$ and of the pair $\left\{y,z\right\}$ are interchanged.

In Fig.~\ref{fig:StdSkewMu}, we depict the coefficient of variation and the skewness of interference for two scenarios; urban  ($c\!=\!4$ m, $r_0\!=\!150$ m) and motorway ($c\!=\!20$ m, $r_0\!=\!500$ m) cells. We calculate the standard deviation as  $\sqrt{\mathbb{E}\!\left\{\mathcal{I}^2\right\}\!-\!\mathbb{E}\!\left\{\mathcal{I}\right\}^2}$, and the skewness as  $\frac{\mathbb{E}\left\{\mathcal{I}^3\right\}-3\mathbb{E}\left\{\mathcal{I}\right\}\mathbb{E}\left\{\mathcal{I}^2\right\}+2\mathbb{E}\left\{\mathcal{I}\right\}^3}{\left(\mathbb{E}\left\{\mathcal{I}^2\right\}-\mathbb{E}\left\{\mathcal{I}\right\}^2 \right)^{3/2}}$, where the terms $\mathbb{E}\!\left\{\mathcal{I}^2\right\},\mathbb{E}\!\left\{\mathcal{I}^3\right\}$ in~\eqref{eq:EI2} and~\eqref{eq:EI3} are evaluated numerically using the  approximations for the terms $S, S', S''$ with $m\!=\!2$. We depict the results up to $\mu\!=\!0.1 {\text{m}}^{-1}$. For $c\!=\!20 $m and $\mu\!=\!0.1 {\text{m}}^{-1}$, we have $\lambda c\!=\!\frac{2}{3}$. For larger $\mu$, the approximation accuracy with $m\!=\!2$ is poor in the motorway scenario because of long-range correlations. The mean and the variance of interference increase for a lower pathloss exponent given all other parameters remain fixed. We see in Fig.~\ref{fig:StdSkewMu} that the coefficient of variation, defined as the ratio of the standard deviation over the mean, becomes smaller. Lower pathloss exponents are associated not only with more concentrated but also with more symmetric, less skewed, interference distributions. Given the pathloss model, the large cell size and tracking distance associated with the motorway scenario have the same effect on the distribution of interference. The distribution becomes more concentrated around the mean and also more symmetric between the tails in comparison with the interference distribution associated with urban microcells.

In Fig.~\ref{fig:StdSkew} we have simulated the standard deviation and the skewness of interference with respect to the tracking distance $c$, while the intensity of vehicles $\lambda$ is fixed. We have used cell sizes, $r_0\!=\!100$ m and $r_0\!=\!150$ m, and pathloss exponents, $\eta\!=\!2$ and $\eta\!=\!3$. We see that the approximations for the \acp{PCF} introduce negligible errors for $\lambda c\!\leq\!0.6$. The approximation for the skewness is more prone to errors because the third moment consists of many terms involving the \ac{PCF} and also, one term with the product of \acp{PCF}. 

Based on Fig.~\ref{fig:StdSkewMu} and Fig.~\ref{fig:StdSkew} we deduce that the Gaussian model for the interference distribution would not be accurate in our system set-up. The distribution is skewed. This is in accordance with the study in~\cite{Ghasemi2008}, illustrating that a two-dimensional \ac{PPP} with a guard zone around the receiver generates a positively skewed \ac{PDF} for the aggregate interference under independent log-normal shadowing. We see in Fig.~\ref{fig:StdSkew} that larger tracking distances make the variance of interference less for a fixed intensity of vehicles. This is intuitive because the deployment becomes more regular. The behaviour of the skewness does not appear to be monotonic.  

The approximations we got so far do not provide much insight into the behaviour of second and third moment of interference, due to the complex nature of $S, S', S''$. We would like to capture the impact of tracking distance on the moments of interference using a simple expression. Assuming an intensity $\lambda$ of vehicles, how do the moments due to a hardcore process, $c>0$, scale as compared to the respective moments due to a \ac{PPP} of equal intensity $\lambda$? Next, we assume, in addition, a small tracking distance $c$ as compared to the cell size $r_0$. Under small $\lambda c$ and $\frac{c}{r_0}$, we will relate the standard deviation of interference to that due to a \ac{PPP}, and draw useful remarks. The more complicated study about the behaviour of skewness, and the selection of appropriate models to describe the \ac{PDF} of interference are left for future work. 

\section{Closed-form approximation for the variance}
\label{sec:VarApprox}
The contribution to the second moment of interference due to pairs of vehicles at distances larger than $2c$ is given by the second term in equation~\eqref{eq:Iapprox}. Let us denote it by $S_{>2c}$. After substituting the propagation pathloss function we get 
\begin{equation}
\label{eq:G00}
\begin{array}{ccl}
S_{>2c} \!\!\!\!\!\!\!& = &\!\!\!\!\!\!\! \displaystyle 2\lambda^2\!\!\left(\int\limits_{r_0}^\infty\!\!\int\limits_{x+2c}^\infty\!\!\!\!\! x^{\!-\eta} y^{\!-\eta} {\rm d}y {\rm d}x  \!+\!\!\! \int\limits_{r_0}^\infty\!\int\limits_{-\infty}^{x-2c}\!\!\!\!\! x^{\!-\eta} g\!\left(y\right)  {\rm d}y {\rm d}x \!\right) \\ 
\!\!\!\!\!& \stackrel{(a)}{=}&\!\!\!\!\! \displaystyle \frac{1}{2}\mathbb{E}\!\left\{\mathcal{I}\right\}^{\!2} \!\!+\! 2\lambda^2\!\! \left(\int\limits_{r_0}^\infty\!\int\limits_{x+2c}^\infty\!\!\!\! \frac{{\rm d}y {\rm d}x}{x^{\eta} y^{\eta}}  \!+\!\!\!\!\! \int\limits_{r_0+2c}^\infty\!\!\!\int\limits_{r_0}^{x-2c}\!\!\!\! \frac{{\rm d}y {\rm d}x}{x^{\eta} y^{\eta}} \right) \\
\!\!\!&\stackrel{(b)}{=}&\!\!\! \displaystyle \frac{1}{2}\mathbb{E}\!\left\{\mathcal{I}\right\}^2 \!\!+\! 4\lambda^2 \!\!  \int_{r_0}^\infty\!\int_{x+2c}^\infty\!\!\! x^{-\eta} y^{-\eta} {\rm d}y {\rm d}x \\ 
\!\!\!&\stackrel{(c)}{=}&\!\!\! \displaystyle \frac{1}{2}\mathbb{E}\!\left\{\mathcal{I}\right\}^2 + \frac{2 \lambda^2 r_0^{2-2\eta}}{\eta\!-\! 1} \!\bigg(  \frac{\left(2b+1\right)^{1-\eta}}{\eta\!-\!1} \,\,\, + \\ \!\!\! & & \,\,\, \displaystyle \frac{2b}{2\eta\!-\!1} {}_2F_1\!\left(\eta,2\eta\!-\!1,2\eta;-2b\right) \bigg), 
\end{array}
\end{equation}
where $(a)$ follows from $2\lambda^2\int_{r_0}^\infty\!\int_{-\infty}^{-r_0}\! x^{-\eta} \left|y\right|^{-\eta}{\rm d}y {\rm d}x\!=\! \frac{1}{2}\mathbb{E}\!\left\{\mathcal{I}\right\}^2$, $(b)$ from symmetry, in $(c)$ we substitute $b=\frac{c}{r_0}$, and ${}_2F_1$ is the Gaussian hypergeometric function~\cite[pp.~556]{Abramo}. 

Let us denote by $S_{<2c}$ the first term of $S$ in~\eqref{eq:Iapprox}, i.e., the contribution to the second moment from pairs of vehicles at distance separation less than $2c$. Due to the common pathloss function $g$ over the users, the contributions to $S_{<2c}$ for $y\!>\!x$ and $x\!<\!y$ are equal (for $c\!<\!r_0$), and thus 
\begin{equation}
\label{eq:G10}
S_{<2c} = 4\lambda\mu\!\int_{r_0}^\infty\!\int_{x+c}^{x+2c}\!\!\!\!\!\!\! x^{-\eta} y^{-\eta} e^{-\mu\left(y-x-c\right)}{\rm d}y {\rm d}x. 
\end{equation}

\noindent 
After integrating in terms of $y$ we have 
\begin{equation}
\begin{array}{ccl}
S_{<2c} \!\!\!\!\!\!\!\!& = &\!\!\!\!\!\! \displaystyle \frac{4\lambda}{\mu^{-\eta}}\!\!\! \int\limits_{r_0}^\infty\! \frac{\Gamma\!\left(1\!-\!\eta,\!\left(c\!+\!x\right)\!\mu\right)  \!-\!  \Gamma\!\left(1\!-\!\eta,\!\left(2c\!+\!x\right)\!\mu\right)}{x^{\eta} e^{-\mu\left(c+x\right)}} {\rm d}x\!, 
\end{array}
\end{equation}
where $\Gamma\!\left(a,x\right)\!=\!\int_x^\infty \! \frac{t^{a-1}}{e^t} {\rm d}t$ is the incomplete Gamma function.

We cannot express the above integral in terms of well-known functions. In order to approximate it, we expand the integrand around  $\mu\left(x\!+\!c\right)\!\rightarrow\!\infty$. For a fixed $\lambda$ and $c\!>\!0$, we have $\mu\!=\!\frac{\lambda}{1-\lambda c}\!>\!\lambda$. In addition, $\left(x\!+\!c\right)\!>\!r_0$. Therefore the expansion should be valid for $\lambda r_0\!\gg\! 1$, i.e., the average number of vehicles within the cell must be  high. After expanding up to the first-order term and carrying out the integration we get 
\begin{equation}
\label{eq:G1}
\begin{array}{ccl}
S_{<2c} \!\!\!\!\!\!&\approx &\!\!\!\!\! \displaystyle  \int\limits_{r_0}^\infty\!\!\!\! \frac{4\lambda}{\left(x\!\left(x\!+\!c\right)\right)^{\!\eta}} \bigg(\!\frac{1\!\!-\!e^{\!-c\mu}}{\mu} \!+\!\! \frac{\eta\!\left(\!e^{\!-c\mu}\!\!\left(1\!\!+\!c\mu\!\right)\!-\!\!1\!\right)}{\mu^2\left(x+c\right)} \!\bigg) \! {\rm d}x \\ \!\!\!\!\!\!&=&\!\!\!\!\! \displaystyle \frac{4\lambda \! \left(1\!-\!e^{-c\mu}\right) {}_2F_1\!\left(\eta,2\eta\!-\!1,2\eta,-\frac{c}{r_0}\right)}{\left(2\eta-1\right) \, r_0^{2\eta-1}} \,\,\, + \\ 
\!\!\!\!\!\!\!\!& &\!\!\!\!\!\!\!\!\!\!\!\!\!\!\!\!  \displaystyle \frac{2\lambda \! \left(e^{\!-c\mu}\!\left(\!1\!+\!c\mu\!\right)\!-\!1\right) {}_2F_1\!\left(\!2\eta,\eta\!+\!1,2\eta\!+\!1,-\frac{c}{r_0}\!\right)}{\mu \, r_0^{2\eta}}. 
\end{array}
\end{equation}
\begin{figure*}[!t]
 \centering
  \subfloat[$r_0=100$ m]{\includegraphics[width=2.5in]{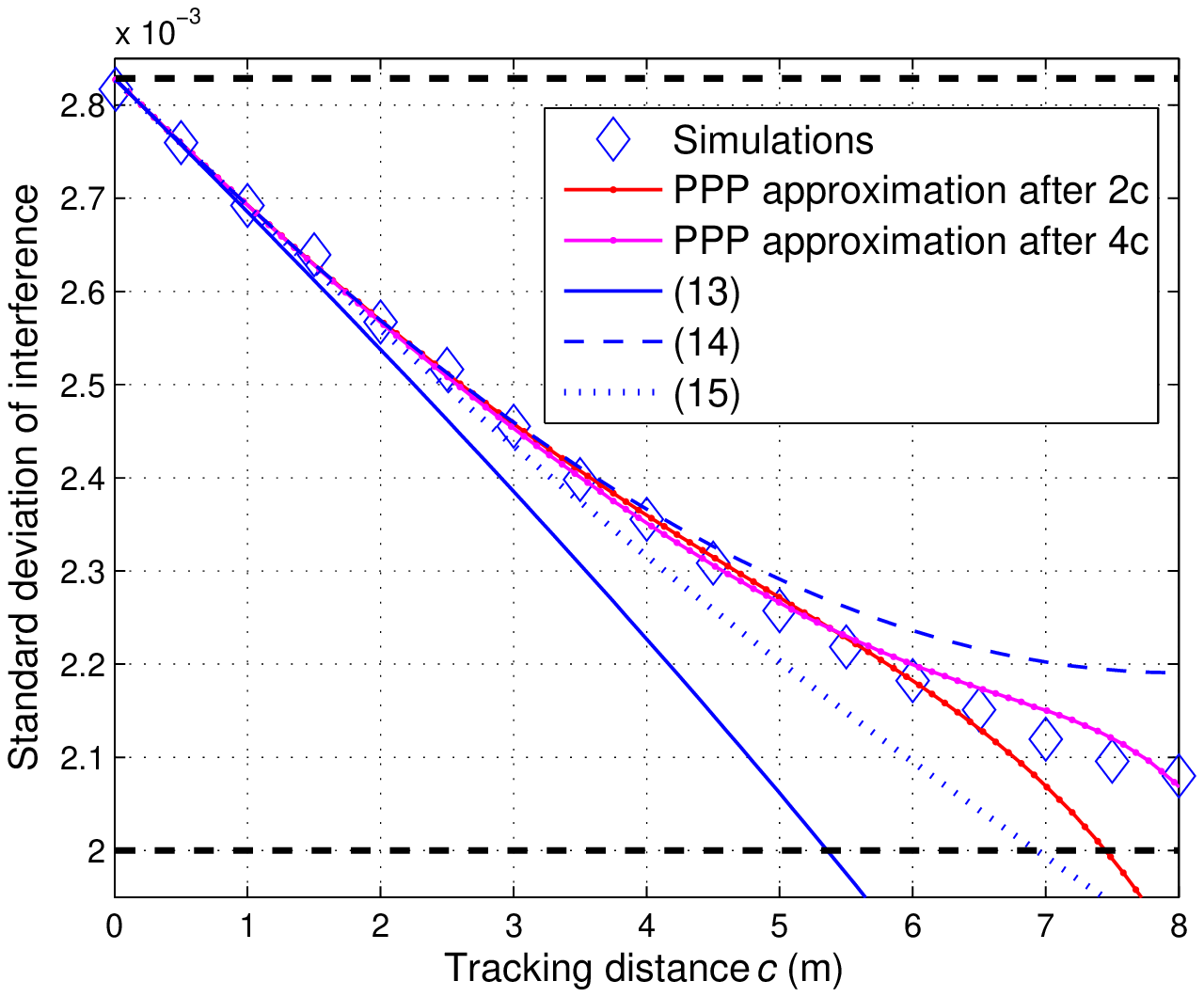}}\hfil \subfloat[$r_0=150$ m]{\includegraphics[width=2.5in]{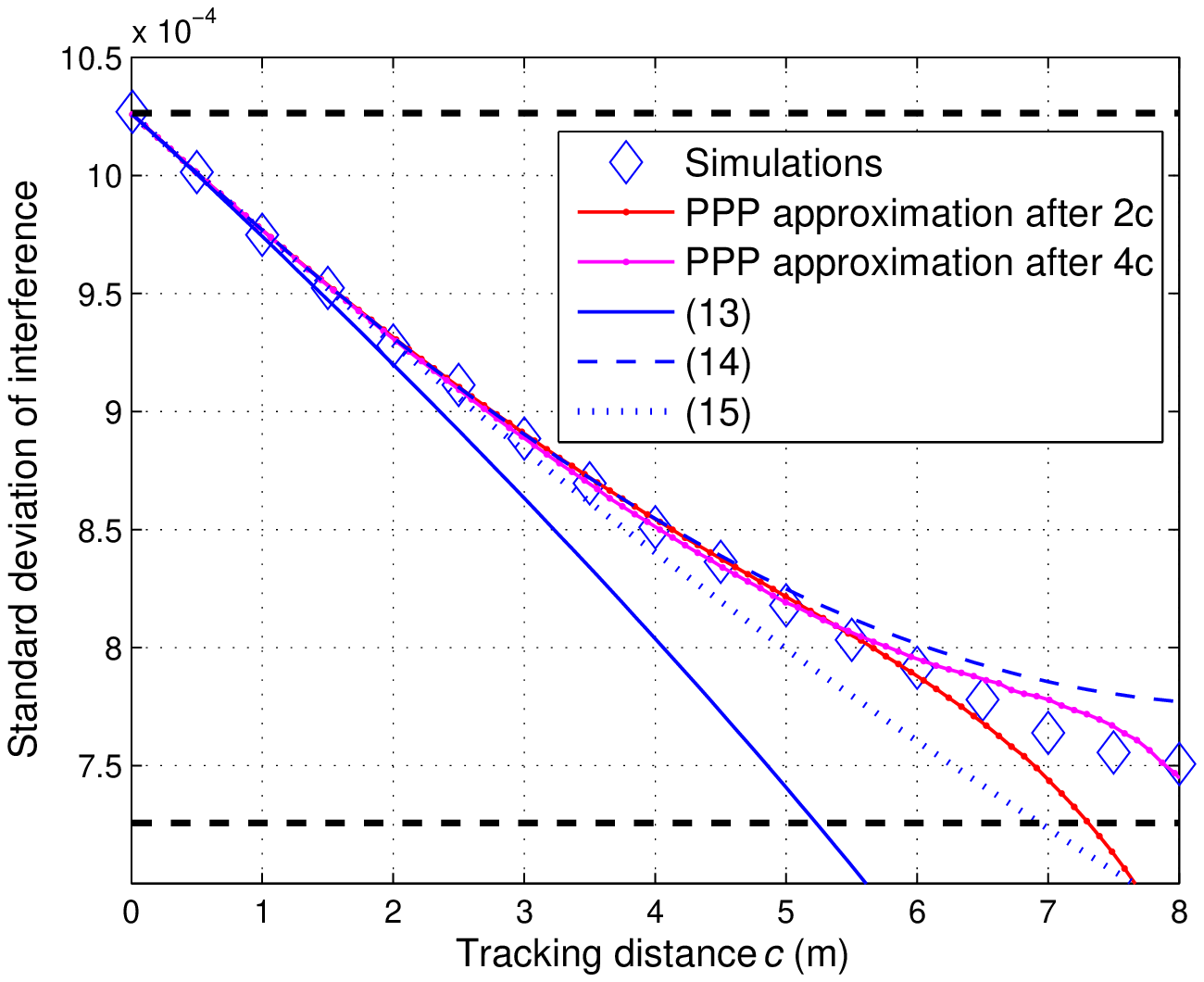}}
 \caption{Standard deviation of interference with respect to the tracking distance. The intensity is $\lambda\!=\! 0.1 {\text{m}}^{-1}$. $2\times 10^5$ simulation runs per marker. Pathloss exponent $\eta\!=\!3$. For the '\ac{PPP} approximation after $2c$' we calculate $S_{>2c}$ from~\eqref{eq:G00}, and $S_{<2c}$ numerically from~\eqref{eq:G10}. In the numerical calculation of '\ac{PPP} approximation after $4c$', similar (but more) integrals to~\eqref{eq:G00} and~\eqref{eq:G10} are involved. The dashed line at the top corresponds to a \ac{PPP} of intensity $\lambda$. The dashed line at the bottom corresponds to a lattice with inter-point distance $\lambda^{-1}$. The details for the calculation of the variance due to a lattice are given in Section~\ref{sec:Lattice}.}
 \label{fig:StdLam1}
\end{figure*}
\begin{figure*}[!t]
 \centering
  \subfloat[$r_0=100$ m]{\includegraphics[width=2.5in]{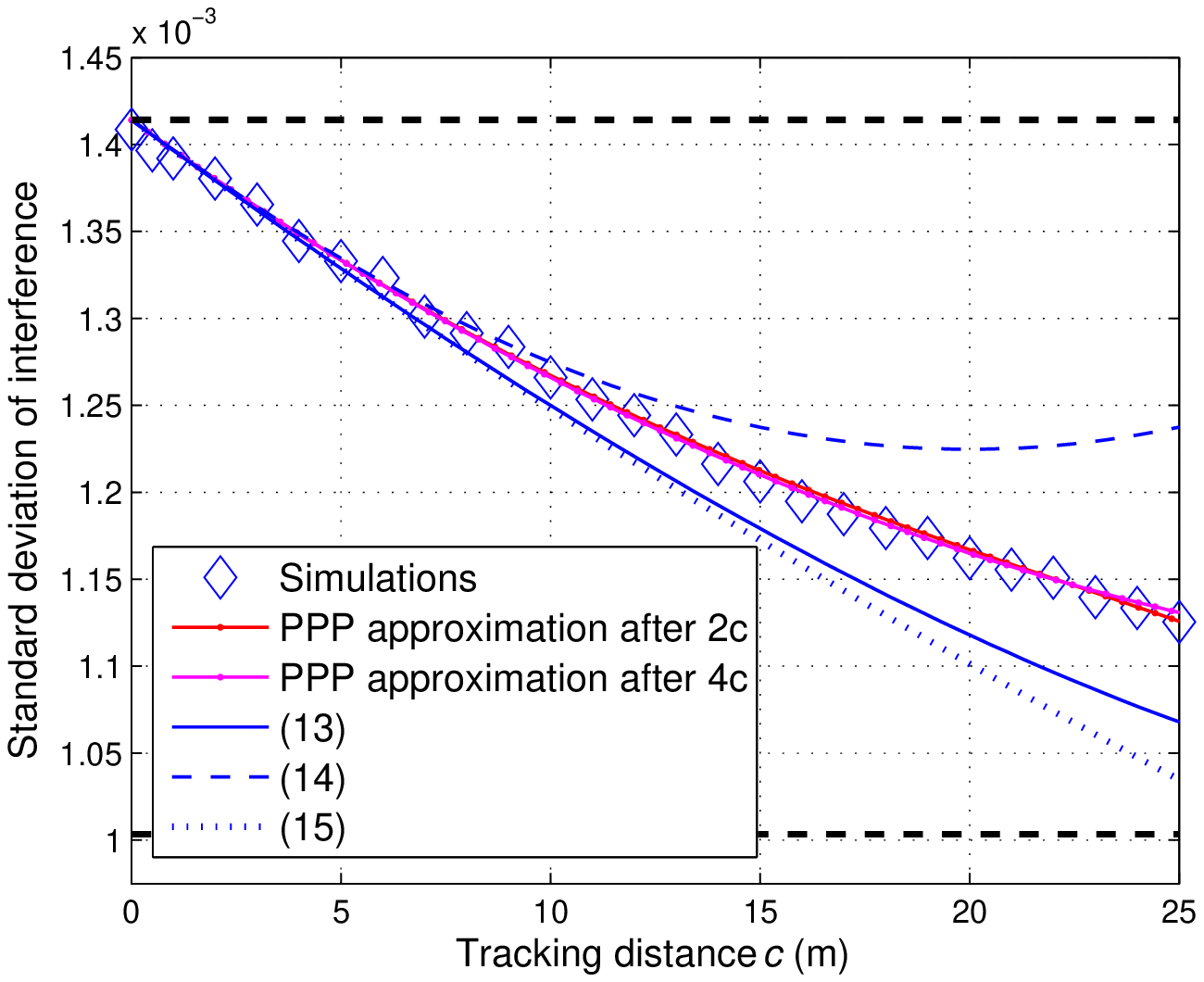}}\hfil \subfloat[$r_0=150$ m]{\includegraphics[width=2.5in]{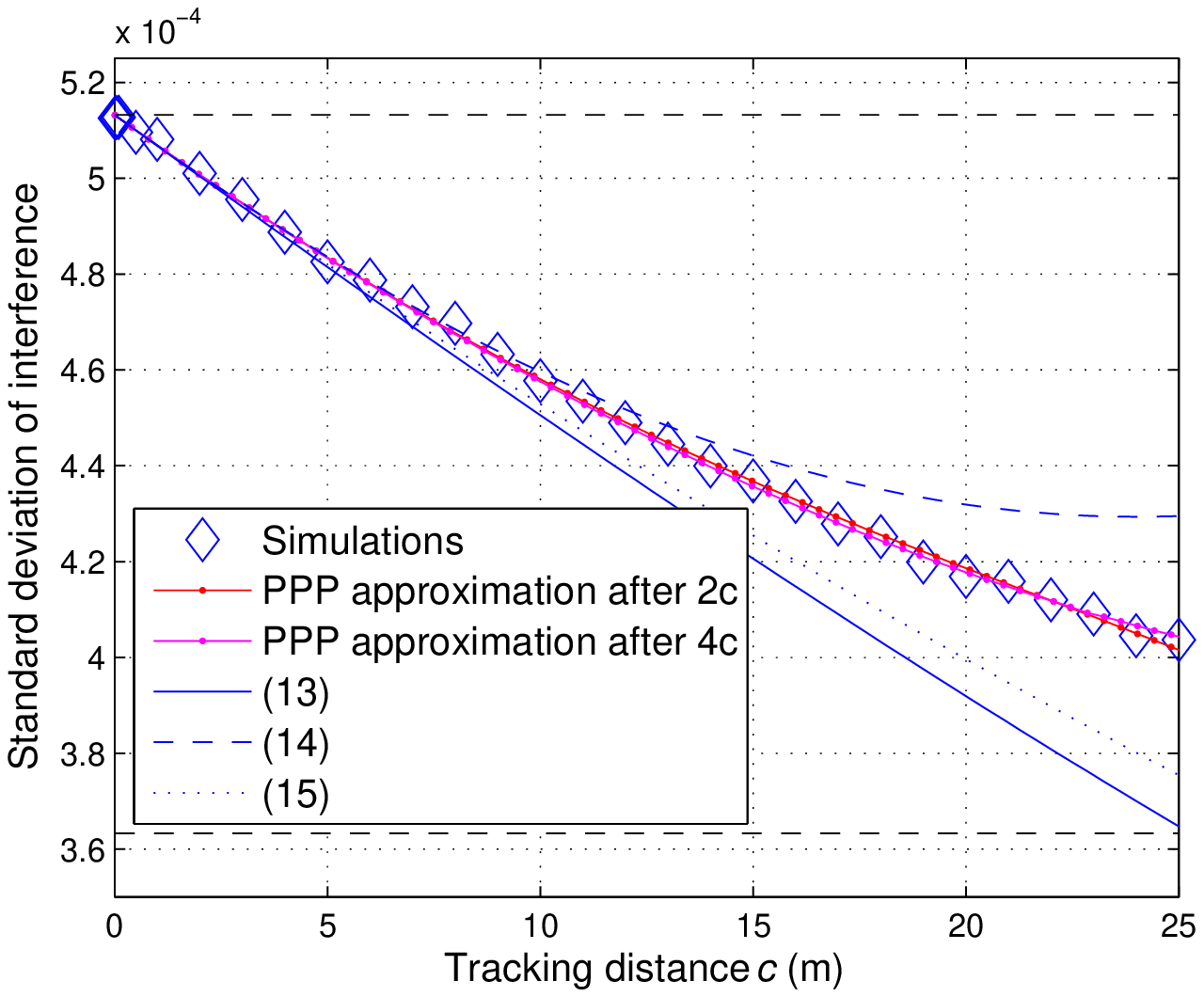}}
 \caption{Standard deviation of interference with respect to the tracking distance $c$. The intensity of vehicles is $\lambda\!=\! 0.025 {\text{m}}^{-1}$. See the caption of Fig.~\ref{fig:StdLam1} for parameter settings and explanation of the legends.}
 \label{fig:StdLam025}
\end{figure*}

For positive $\lambda$, the relation $\lambda c\!=\! 1$ corresponds to a lattice with inter-point distance $c\!=\!\lambda^{-1}$, and the relation $\lambda c \!= \! 0$ corresponds to a \ac{PPP} of intensity $\lambda$. We would like to approximate the variance of interference for small $c$, while $\lambda$ remains fixed. We start from the approximation of the term $S_{<2c}$ in~\eqref{eq:G1}, we substitute $\mu\!=\!\frac{\lambda}{1-\lambda c}$, and expand around $\lambda c\!\rightarrow\!\ 0$, up to second-order. 
\begin{equation}
\label{eq:G12}
\begin{array}{lll}
S_{<2c} \!\!\!&\approx&\!\!\! \displaystyle  2 \lambda^2 c^2 \bigg( \frac{2 r_0^{1-2\eta} {}_2F_1\!\left(2\eta\!-\!1,\eta,2\eta,-\frac{c}{r_0}\right)}{\left(2\eta-1\right) c} \, - \\ \!\!\!& &\!\!\! \displaystyle \frac{r_0^{-2\eta}}{2}{}_2F_1\!\left(2\eta,\eta\!+\!1,2\eta\!+\!1,-\frac{c}{r_0}\right)\! \bigg). 
\end{array}
\end{equation}

After substituting~\eqref{eq:Iapprox} into~\eqref{eq:EI2}, noting that $S\!=\!S_{>2c}\!+\!S_{<2c}$, carrying out the integration describing the contribution to the second moment from a single vehicle, we get 
\begin{equation} 
\label{eq:VarApprox}
\mathbb{V}{\text{ar}}\left\{\mathcal{I}\right\} \approx \frac{4 \lambda r_0^{1-2\eta}}{2\eta-1} \!+\! S_{>2c}\!+\! S_{<2c} \!-\! \mathbb{E}\!\left\{\mathcal{I}\right\}^2.
\end{equation}

\noindent 
Next, we substitute~\eqref{eq:G00} and~\eqref{eq:G12} in~\eqref{eq:VarApprox}. 
\[
\begin{array}{ccl}
\mathbb{V}{\text{ar}}\left\{\mathcal{I}\right\} \!\!\!\!\!&\approx&\!\!\!\!\! \displaystyle \frac{4 \lambda r_0^{1-2\eta}}{2\eta-1}  + \frac{2 \lambda^2 r_0^{2-2\eta}}{\eta\!-\! 1} \!\bigg( \frac{\left(2b+1\right)^{1-\eta}}{\eta\!-\!1} \, + \\ \!\!\!\!\!\!\!\!\!\!& &\!\!\!\!\!\!\!\!\!\! \displaystyle \frac{2b\,{}_2F_1\!\left( \eta,2\eta\!-\!1,2\eta;-2b\right)}{2\eta\!-\!1} \bigg) \!-\! \frac{\mathbb{E}\!\left\{\mathcal{I}\right\}^2}{2} \!+\! 2\lambda^2 r_0^{2-2\eta} \times \\ \!\!\!\!\!\!\!\!\!\!\!\!\!\!\!\!\!\!\!\!\!\!\!\!\!& &\!\!\!\!\!\!\!\!\!\!\!\!\!\!\!\!\!\!\!\!\!\!\!\!\! \displaystyle \left( \frac{2 b\,   {}_2F_1\!\left(2\eta\!-\!1,\!\eta,\!2\eta,\!-b\right)}{2\eta-1} -\frac{b^2 {}_2F_1\!\left(2\eta,\!\eta\!+\!1,\!2\eta\!+\!1,\!-b\right)}{2} \right)\!. 
\end{array}
\]

\noindent 
After expanding up to second order in $b\!\rightarrow\! 0$ we have 
\begin{equation} 
\label{eq:VarApprox2}
\mathbb{V}{\text{ar}}\left\{\mathcal{I}\right\} \approx \frac{4 \lambda r_0^{1-2\eta}}{2\eta\!-\!1}\! \left(1\!-\!\lambda c \right)  \!+\! \lambda^2 c^2  r_0^{-2\eta}.  
\end{equation}

If we approximate the \ac{PCF} one step further ($m\!=\!3$ instead of $m\!=\!2$), and repeat the same procedure, we end up with  
\begin{equation} 
\label{eq:VarApprox4}
\mathbb{V}{\text{ar}}\left\{\mathcal{I}\right\} \approx \frac{4 \lambda r_0^{1-2\eta}}{2\eta\!-\!1}\! \left(1\!-\!\lambda c \!+\!\frac{\lambda^2 c^2}{2}\right)  \!+\! \lambda^2 c^2  r_0^{-2\eta}.  
\end{equation}

The leading order term, $r_0^{1-2\eta}$, in~\eqref{eq:VarApprox2} and~\eqref{eq:VarApprox4}, will dominate the variance for $r_0\gg c$. In addition, for small $\lambda c$, we can use the expansion of the exponential function around zero, $e^{-\lambda c} \!\approx\! 1\!-\!\lambda c\!+\!\frac{\lambda^2 c^2}{2}$, to get 
\begin{equation} 
\label{eq:VarApprox3}
\mathbb{V}{\text{ar}}\left\{\mathcal{I}\right\} \approx \frac{4 \lambda r_0^{1-2\eta}}{2\eta\!-\!1} e^{-\lambda c}. 
\end{equation}

The above approximation relates in a simple manner the variance of interference due to a \ac{PPP} of intensity $\lambda$, with the variance of interference due to a hardcore process of equal intensity, for small $\lambda c$. Introducing a tracking distance $c$, while keeping the intensity $\lambda$ fixed, makes the deployment more regular, and this results in exponential reduction $e^{-\lambda c}$ for the variance of interference, or equivalently, $e^{-\frac{\lambda c}{2}}$, for the standard deviation. The linear reduction (in logarithmic scale) of the standard deviation with respect to $c$ is evident in Fig.~\ref{fig:StdSkew}. Using the approximation for the variance in~\eqref{eq:VarApprox3}, the coefficient of variation can be read as $\frac{\eta-1}{\sqrt{2\eta-1}} \frac{1}{\sqrt{\lambda r_0}}\, e^{-\frac{\lambda c}{2}}$. This approximation for the correlation coefficient agrees with the illustrations by Fig.~\ref{fig:StdMu}, and allows us to draw the following conclusion: The distribution of interference becomes more concentrated around the mean for smaller pathloss exponent $\eta$, larger cell size $r_0$, and increasing hardcore distance $c$ while $\lambda$ remains fixed. 

In Fig.~\ref{fig:StdLam1}, we have simulated the standard deviation of interference for high traffic conditions $\lambda\!=\! 0.1 {\text{m}^{-1}}$, i.e., on average one vehicle per $10$ m. We depict the results for $\lambda c\in\left(0,0.8\right)$. We see that the closed-form models~\eqref{eq:VarApprox2}$-$\eqref{eq:VarApprox3} are indeed valid for small $c$. The model in~\eqref{eq:VarApprox4} provides a good fit also for realistic tracking distances. This is because it uses the exact \ac{PCF} up to $3c$ instead of $2c$. The considered distances, $c\in\left(0,8\right)$ m are much smaller than the cell size $r_0$, thereby the expansions around $b\!\rightarrow\! 0$ are accurate too. For tracking distances $c\!>\! 6$ m, the model using the exact \ac{PCF} only up to $2c$ starts to fail, because of larger range correlations.  

In Fig.~\ref{fig:StdLam025}, we replicate the results of Fig.~\ref{fig:StdLam1} for lower traffic intensity, on average, one vehicle per $40$ m. The average inter-vehicle distance becomes comparable to the cell size $r_0$, and the feasible tracking distances span a much larger range. We depict the results up to $c\!=\! 25$ m, or equivalently $\lambda c\in \left(0,0.625\right)$. We deduce that the models~\eqref{eq:VarApprox2}$-$\eqref{eq:VarApprox3} do not fail due to the approximation of the \ac{PCF}. We also note that the source of error is the approximation in $b\!\rightarrow\! 0$ rather than the expansion around $\lambda c\!\rightarrow\! 0$. The models~\eqref{eq:VarApprox2}$-$\eqref{eq:VarApprox3} are still valid for small tracking distances $c$. For realistic values of $c$, they give much more accurate predictions than the \ac{PPP}. 

\section{Interference due to a lattice}
\label{sec:Lattice}
\begin{figure}[!t]
 \centering
  \includegraphics[width=3.5in]{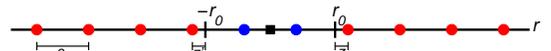}
 \caption{One-dimensional lattice. The base station, 'black square', is located at the origin. Interference is due to points outside of the cell, 'red disks'. The \acp{RV} $z, z'$ represent distances between the cell border and the lattice point nearest to it generating interference.}
 \label{fig:SystModelLattice}
\end{figure}
In the previous section, we constructed simple closed-form models for the variance of interference due to a hardcore process.  These models fail to  describe the variance with long-range correlations, i.e., $\lambda c\!\rightarrow \! 1$. Due to its high complexity, we leave this study for future, and study the extreme scenario, $\lambda c\!=\! 1$, to get a preliminary insight. For $\lambda c\!=\! 1$, the locations of vehicles form a lattice. Studying the moments of interference due to infinite lattices is also a preliminary step before incorporating more complicated deployments in our analysis, e.g., Cowan M3~\cite{Cowan1975}. According to this model, finite lattices of geometrically distributed sizes are separated by exponentially distributed gaps, modeling bunches of vehicles with gaps in-between the bunches.

The performance of lattice networks (not only \ac{1D}) has been studied in~\cite{Haenggi2010}. Over there, the location of the receiver associated to the transmitter at the origin is optimized to maximize the achievable rate. Given the receiver's location, the interference from all points $\mathbb{Z} \!\setminus\! \left\{o\right\}$ becomes deterministic. In our system set-up, the sources of randomness are the Rayleigh fading and the distance $z$ between the cell border $r_0$ and the nearest point to it generating interference, see Fig.~\ref{fig:SystModelLattice}. Since we know the locations of all interferers given $z\!\in\!\left(0,c\right)$, the \ac{PCF} becomes an infinite series of Dirac delta functions, and this will greatly simplify the derivation of higher-order moments. For instance, we will show that the double integration $\int\!\! g\!\left(x\right) g\!\left(y\right) \rho^{\left(2\right)}\!\left(x,y\right) {\rm d}x {\rm d}y$ in the calculation of second moment degenerates to single integration in terms of $z$. The instantaneous interference at the base station due to the lattice points located at the positive half-axis is  
\[
\begin{array}{ccl}
\mathcal{I} = \displaystyle \sum\limits_{k=0}^\infty h_k \, g\!\left(x_k\right) = \sum\limits_{k=0}^\infty h_k \, g\!\left(r_0\!+\!z\!+\!k c\right), 
\end{array}
\]
where $h_k, x_k$ are the fading coefficient and location for the $k$-th point respectively, and $z$ is a uniform \ac{RV}, $z\!=\! U\left(0,c\right)$.

\noindent 
The \ac{MGF} of interference is 
\[
\begin{array}{ccl}
\Phi_{\mathcal{I}}\!\left(s\right)\!=\! \displaystyle \int e^{s\mathcal{I}} f_{\rm h} f_{\rm x} {\rm dh dx}, 
\end{array}
\]
where ${\rm h,x}$ are the vectors of fading coefficients and user locations respectively, and $f_{\rm h}, f_{\rm x}$ are the associated \acp{PDF}. 

The mean interference can be calculated by evaluating the first derivative of the \ac{MGF} at $s\!=\!0$.
\begin{equation}
\label{eq:EILattice}
\begin{array}{ccl}
\mathbb{E}\!\left\{\mathcal{I}\right\} \!\!\!\!\!&=&\!\!\!\!\! \displaystyle \frac{\partial \Phi_{\mathcal{I}}}{\partial s}\Big|_{s=0} = 2\!\int \sum\limits_{k=0}^\infty \!\! h_k  g\left(x_k\right)  f_{\rm h} f_{\rm x} {\rm dh dx} \\  \!\!\!&\stackrel{(a)}{=}&\!\!\! \displaystyle 2\!\int \sum_{k=0}^\infty h_k \, g\!\left(r_0\!+\!z\!+\!kc\right) f_{\rm h} f_z {\rm dh d}z \\ 
\!\!\!& \stackrel{(b)}{=} &\!\!\! \displaystyle 2\!\int \sum\nolimits_{k=0}^\infty g\!\left(r_0\!+\!z\!+\!kc\right) f_z {\rm d}z \\ \!\!\!&=&\!\!\! \displaystyle \frac{2}{c} \int_0^c  \sum\nolimits_{k=0}^\infty \left(r_0\!+\!z\!+\!kc\right)^{-\eta} {\rm d}z \\ \!\!\!&=&\!\!\! \displaystyle \frac{2}{c^{1+\eta}} \int_0^c \zeta\!\left(\eta,\frac{r_0+z}{c}\right) {\rm d}z,
\end{array}
\end{equation}
where the factor two has been added to account for lattice points in the negative half-axis, $(a)$ is due to the fact that given $z$, the locations of  all points become nonrandom, $(b)$ follows from independent fading coefficients and $\mathbb{E}\!\left\{h_k\right\}\!=\!1$, and $\zeta\!\left(n,x\right)\!=\!\sum\nolimits_{k=0}^\infty \left(k+x \right)^{-n}$ is the Hurwitz Zeta function. 

\noindent 
After carrying out the integration in~\eqref{eq:EILattice}, 
\begin{equation}
\label{eq:appEI}
\begin{array}{ccl}
\mathbb{E}\!\left\{\mathcal{I}\right\} \!\!\!\!\!&=&\!\!\!\!\! \displaystyle \frac{2\left( \zeta\!\left(\eta\!-\!1,q\right) \!-\! \zeta\!\left(\eta\!-\!1,1\!+\!q\right) \right)}{c^{\eta}\left(\eta-1\right)} \stackrel{(a)}{=} \frac{2r_0^{1-\eta}}{c\left(\eta\!-\!1\right)}, 
\end{array}
\end{equation}
where $q\!=\!\frac{r_0}{c}$, and in $(a)$ we have used the identity for consecutive neighbors  $\zeta\!\left(n,x\right)=\zeta\!\left(n,1\!+\!x\right)\!+\!x^{-n}$.

Due to the Campbell's Theoreom~\cite{Mecke2013}, the mean interference can also be calculated by averaging the distance-based pathloss over the intensity of lattice points $\mathbb{E}\!\left\{\mathcal{I}\right\} = 2\lambda\int_0^\infty\! \left(r+r_0\right)^{-\eta} {\rm d}r = \frac{2\lambda r_0^{1-\eta}}{\eta-1}$, where the intensity $\lambda\!=\!c^{-1}$. 

In order to calculate the second moment of interference, we need to consider explicitly the interference originated from the negative half-axis. For that we have to identify the conditional \ac{PMF} of the distance $z'$ between the cell border $-r_0$ and the nearest lattice point to it generating interference, given the distance $z$, see Fig.~\ref{fig:SystModelLattice}. Let us denote $\epsilon\!=\frac{2r_0}{c} - \lfloor \frac{2r_0}{c} \rfloor$. The conditional \ac{PMF} becomes equal to $z'\!=\!\left(c\left(1\!-\!\epsilon \right)\!-\!z\right)$ with probability $\left(1\!-\!\epsilon\right)$, and equal to  $z'\!=\!\left(c\left(2\!-\!\epsilon\right)\!-\!z\right)$ with probability $\epsilon$. For presentation clarity, we will assume that the diameter of the cell, $2r_0$, is an integer multiple of the inter-point distance, i.e., $\epsilon\!=\! 0$. In that case, $z'\!=\!\left(c\!-\!z\right)$ with probability one. Extensions and numerical results for a positive $\epsilon$ will be given. 

The second moment of interference can be calculated by evaluating the second  derivative of the \ac{MGF} at $s\!=\!0$. 
\[
\begin{array}{ccl}
\mathbb{E}\!\left\{\mathcal{I}^2\right\} \!\!\!\!\!\!\!&=&\!\!\!\!\!\!\! \displaystyle \frac{\partial^2 \Phi_{\mathcal{I}}}{\partial s^2}\Big|_{s=0} \\ \!\!\!\!\!\!\!&=&\!\!\!\!\!\!\! \displaystyle \int\! \left( \sum_{k=0}^\infty  h_k\,  g\!\left(x_k\right) \!+\! \sum_{m=0}^\infty  h_m\,  g\!\left(x_m\right) \right)^{\!\!\!2} \! f_{\rm h} f_{\rm x} {\rm dh dx} \\ \!\!\!\!\!\!\!&=&\!\!\!\!\!\!\! \displaystyle \int \!\!\! \left(\! 2\!\! \left( \sum\limits_{k=0}^\infty  \!\! h_k g\!\left(\!x_k\!\right)\!\!\right)^{\!2}  \!\!\!\!+\!\!
2\!\sum_{k,m}\!\!  h_k h_m  g\!\left(\!x_k\!\right) \! g\!\left(\!x_m\!\right) \!\! \right) \!\!\! f_{\rm h} \! f_{\rm x}  {\rm dh dx}, 
\end{array}
\]
where the sum over $m$ describes the interference from the negative half-axis, and the factor two in front of the square term is due to symmetry.

\noindent
After expanding the square term we have 
\[
\begin{array}{ccl}
\mathbb{E}\!\left\{\mathcal{I}^2\right\} \!\!\!\!\!\!\!\!&=&\!\!\!\!\!\!\!\! \displaystyle  \int \!\bigg(\!2 \Big(\sum_{k=0}^\infty  \! h_k^2  g\!\left(x_k\right)^2 \!\!+\!\!  \sum_{k=0}^\infty\sum_{k'\neq k} \! h_k h_{k'}  g\!\left(x_k\right) g\!\left(x_{k'}\right)\!  \Big) \!+ \\ & &\ \,\,\,\,\,\,\, \displaystyle 2\sum_{k=0}^\infty\sum_{m=0}^\infty  h_k h_m g\!\left(x_k\right) g\!\left(x_m\right) f_{\rm h} f_{\rm x}\bigg) {\rm dh dx} \\ \!\!\!\!\!\!\!&\stackrel{(a)}{=}&\!\!\!\!\!\!\! \displaystyle  \int \!\bigg(2  \Big(\!\sum_{k=0}^\infty 2  g\!\left(x_k\right)^2 \!+\!  \sum_{k=0}^\infty\sum_{k'\neq k} g\!\left(x_k\right) g\!\left(x_{k'}\right)\! \Big)  \, + \\ {} & & \,\,\,  \displaystyle  2\sum_{k=0}^\infty\sum_{m=0}^\infty g\!\left(x_k\right) g\!\left(x_m\right) \bigg) f_{\rm x} {\rm dx} \\ 
\!\!\!\!\!\!\!\!&\stackrel{(b)}{=}&\!\!\!\!\!\! \displaystyle \underbrace{2\!\! \int \sum_{k=0}^\infty   \! g\!\left(x_k\right)^{\!2} \! f_{\rm x} {\rm dx}}_{J_1} \!+\! \underbrace{2\!\! \int\! \sum_{k=0}^\infty\sum_{k'=0}^\infty \!g\!\left(x_k\right) g\!\left(x_{k'}\right) \! f_{\rm x} {\rm dx}}_{J_2} \! + \\ & & \,\,\, \displaystyle \underbrace{2\! \int \sum_{k=0}^\infty\sum_{m=0}^\infty   g\!\left(x_k\right) g\!\left(x_m\right)  f_{\rm x} {\rm dx}}_{J_3},
\end{array}
\]
where $(a)$ is due to $\mathbb{E}\!\left\{h_k^2\right\}\!=\!2$, $\mathbb{E}\!\left\{h_k\right\}\!=\!1$, and independent fading among the users, and in $(b)$ we have added $k'\!=\!k$ in the second sum (so that the sum over $k'$ goes over all positive integers similar to $k$) and subtract it from the first sum. 

The distances $z,z'$ to the cell borders are in general unequal $z\!\neq\!z'$. Therefore $J_2\!\neq\!J_3$ ($k'$ goes over the positive half- while $m$ spans the negative half-axis). The term $J_1$ can be calculated as in equation~\eqref{eq:appEI}, i.e., conditioning in terms of $z$, integrating the Zeta function and using its consecutive neighbors identity
\begin{equation}
\label{eq:J1}
\begin{array}{ccl}
J_1 \!\!\!\!\!\!&=&\!\!\!\!\!\! \displaystyle 2\!\int \sum_{k=0}^\infty g\!\left(r_0\!+\!z\!+\!kc\right)^2  f_z {\rm d}z \\ \!\!\!\!\!\!&=&\!\!\!\!\!\! \displaystyle \frac{2}{c} \int_0^c  \sum_{k=0}^\infty \left(r_0\!+\!z\!+\!kc\right)^{-2\eta} {\rm d}z \\ \!\!\!\!\!\!&=&\!\!\!\!\!\! \displaystyle \frac{2\left( \zeta\!\left(2\eta\!-\!1,q\right) - \zeta\!\left(2\eta\!-\!1,1+q\right) \right)}{c^{2\eta}\left(2\eta-1\right)} = \frac{2\lambda r_0^{1-2\eta}}{2\eta-1}.
\end{array}
\end{equation}

\noindent 
In a similar manner, the terms $J_2$ and $J_3$ can be expressed as  
\begin{equation}
\label{eq:J23}
\begin{array}{ccl}
J_2 \!\!\!\!&=&\!\!\!\! \displaystyle \frac{2}{c}\!\!\int_0^c\!\! \sum_{k=0}^\infty \sum_{k'=0}^\infty \!\left(r_0\!+\!z\!+\!kc\right)^{\!-\!\eta} \left(r_0\!+\!z\!+\!k'c\right)^{\!-\!\eta} \! {\rm d}z \\ \!\!\!\!&=&\!\!\!\! \displaystyle 2c^{-2\eta-1}\int_0^c \zeta\!\left(\eta,\!\frac{r_0\!+\!z}{c}\right)^{\!2} \! {\rm d}z. \\ 
J_3 \!\!\!\!&=&\!\!\!\! \displaystyle \frac{2}{c}\!\!\int_0^c\!\! \sum_{k=0}^\infty\! \sum_{m=0}^\infty \!\!\! \left(r_0\!+\!\!z\!+\!kc\right)^{\!-\!\eta}\! \left(r_0\!+\!c\!-\!\!z\!+\!mc\right)^{\!-\!\eta}\! {\rm d}z \\ \!\!\!\!&=&\!\!\!\! \displaystyle 2c^{-2\eta-1}\!\!\! \int_0^c \!  \zeta\!\left(\eta,\!\frac{r_0\!+\!z}{c}\right) \zeta\!\left(\eta,\!\frac{r_0\!+\!c\!-\!z}{c}\right)\! {\rm d}z. 
\end{array}
\end{equation}

For positive $\epsilon$, the calculation of $J_1, J_2$ and $J_3$  requires to average over the \ac{PMF} of $z'$. The terms $J_1$ and $J_2$ would also include integrals of sums over the negative half-axis; instead of scaling by two the corresponding integrals over the positive half-axis. Due to the fact that the \ac{RV} $z'$ is also uniform, $z'\!=\!U\left(0,c\right)$, the terms $J_1$ and $J_2$ for $\epsilon\!>\! 0$ ends up equal to equations~\eqref{eq:J1} and~\eqref{eq:J23}. The term $J_3$ contains the cross-terms, over the two axes, thus it requires to average over the \ac{PMF} of the \ac{RV} $z'$ given $z$. The term $J_3$ for $\epsilon\!>\! 0$  will be larger than that in~\eqref{eq:J23}, reflecting the extra randomness introduced by the conditional \ac{PMF}. Recall that for an  arbitrary  $\epsilon, z'\!=\!\left(\left(1\!-\!\epsilon\right)c\!-\!z\right)$ for $z\!\leq\!\left(1\!-\!\epsilon\right)c$ and $z'\!=\!\left(\left(2\!-\!\epsilon\right)c\!-\!z\right)$ for $z\!>\!\left(1\!-\!\epsilon\right)c$. Therefore the term $J_3$ becomes 
\begin{equation}
\label{eq:J3eps}
\begin{array}{ccl}
J_3 \!\!\!\!\!&=&\!\!\!\!\! \displaystyle \frac{2}{c^{2\eta+1}} \!\!\!\!\!\!\! \int\limits_0^{\left(1-\epsilon\right)c} \!\!\!\!\!\! \zeta\!\left(\!\eta,\!\frac{r_0\!+\!z}{c}\!\right)  \! \zeta\!\left(\!\eta,\!\frac{r_0\!+\!c\left(1\!-\!\epsilon\right)\!-\!z}{c}\!\right) \! {\rm d}z  +  \\ 
\!\!\!\!\!& &\!\!\!\!\! \displaystyle \frac{2}{c^{2\eta+1}} \!\!\!\!\!\!\! \int\limits_{\left(1-\epsilon\right)c}^c  \!\!\!\!\! \zeta\!\left(\!\eta,\frac{r_0\!+\!z}{c}\!\right)\! \zeta\!\left(\!\eta,\frac{r_0\!+\!c\left(2\!-\!\epsilon\right)\!-\!z}{c}\!\right)  \! {\rm d}z.
\end{array}
\end{equation} 

For $\epsilon\!=\! 0$, equation~\eqref{eq:J3eps} degenerates to the expression of $J_3$ in~\eqref{eq:J23}. Finally, the variance of interference can be read as 
\begin{equation}
\label{eq:VarGrid}
\mathbb{V}{\text{ar}}\left\{\mathcal{I}\right\} = \frac{2\lambda r_0^{1-2\eta}}{2\eta-1} + J_2 + J_3 -\left(\frac{2\lambda r_0^{1-\eta}}{\eta-1}\right)^2. 
\end{equation}

\begin{figure*}[!t]
 \centering
  \subfloat[$r_0=100$ m]{\includegraphics[width=2.5in]{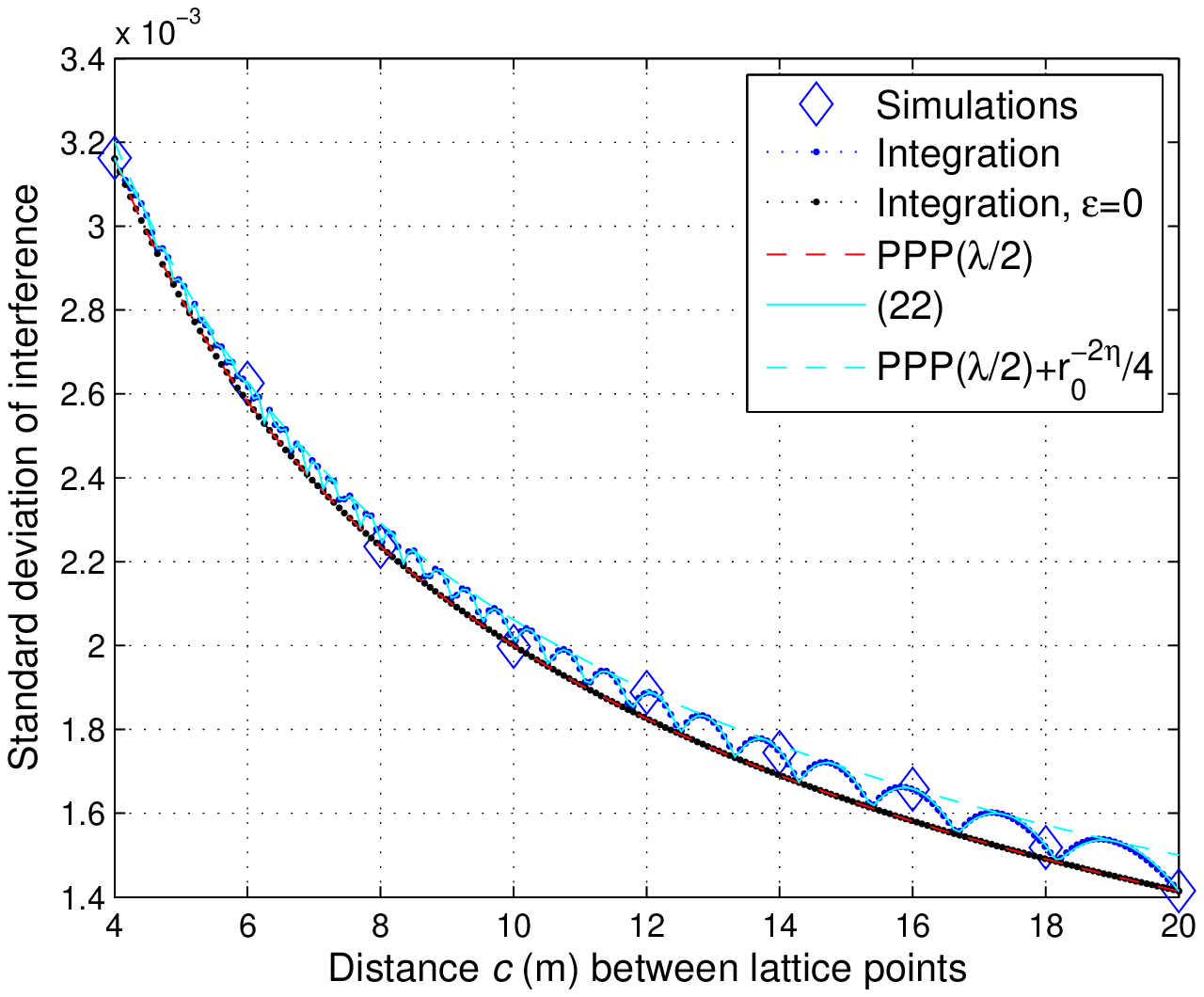}}\hfil \subfloat[$r_0=150$ m]{\includegraphics[width=2.5in]{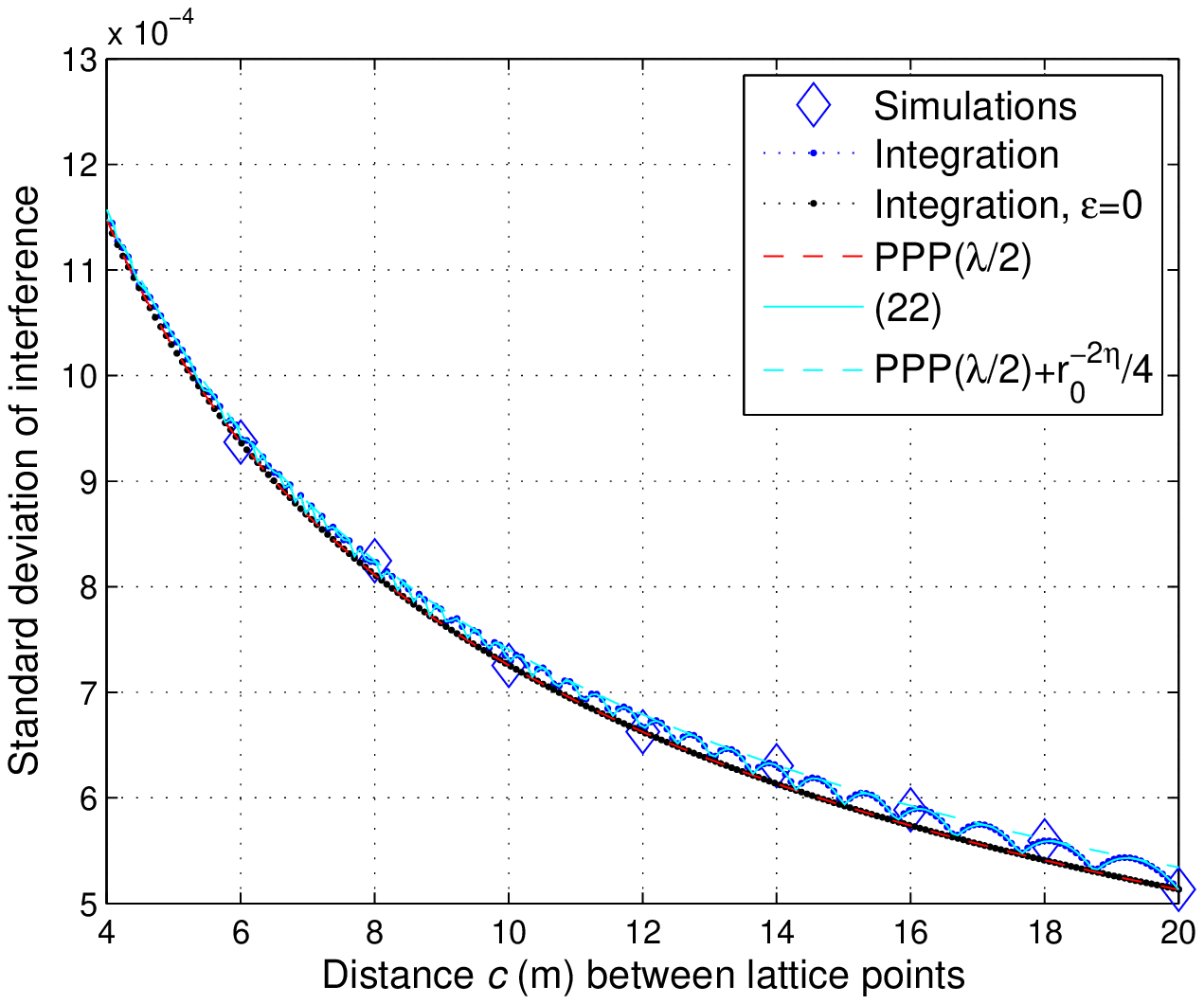}}
 \caption{Standard deviation of interference originated from a lattice. $5\times 10^6$ simulation runs per marker. Pathloss exponent $\eta\!=\!3$. The integration corresponds to equation~\eqref{eq:VarGrid}, where the terms $J_2$ in~\eqref{eq:J23} and $J_3$ in~\eqref{eq:J3eps} are evaluated numerically. The integration with $\epsilon\!=\! 0$ calculates $J_3$ numerically from~\eqref{eq:J23}. The standard deviation of interference due to a \ac{PPP} of intensity $\frac{\lambda}{2}$ is $\sqrt{\frac{2\lambda}{2\eta-1}r_0^{1-2\eta}}$, where $\lambda\!=\! c^{-1}$.}
 \label{fig:StdGrid}
\end{figure*}

In Fig.~\ref{fig:StdGrid}, the integral-based calculation of the variance, see~\eqref{eq:VarGrid} with the term $J_3$ calculated in~\eqref{eq:J3eps}, is verified with the simulations. We include also the calculations with the term $J_3$ calculated in~\eqref{eq:J23}, i.e., $\epsilon\!=\! 0 \,  \forall\left\{c,r_0\right\}$. The impact of positive $\epsilon$ becomes more prominent for higher inter-point distance $c$ while keeping the cell size $r_0$ fixed. 

The terms $J_2$, $J_3$ are difficult to express in closed-form, see~\cite{Shpot2016} for some recent work involving integrals of products of Zeta functions. A high precision evaluation of the Hurwitz Zeta function is also an issue because the function is an infinite sum~\cite{Fredrik}. In order to derive a closed-form approximation for~\eqref{eq:VarGrid}, we note that for large $q\!=\!\frac{r_0}{c}, \epsilon\!=\! 0$ and Rayleigh fading, the variance of interference due to a lattice can be well-approximated by $\frac{2\lambda r_0^{1-2\eta}}{2\eta-1}$. This is because the variance due to a \ac{PPP} under Rayleigh fading, $\frac{4\lambda r_0^{1-2\eta}}{2\eta-1}$, accepts equal contributions, $\frac{2\lambda r_0^{1-2\eta}}{2\eta-1}$, due to fading and due to random user locations. The variance of interference due to a lattice with inter-point distance much less than the cell radius should be random mostly due to the fading, i.e., $\frac{2\lambda r_0^{1-2\eta}}{2\eta-1}$. In Fig.~\ref{fig:StdGrid}, we see that the corresponding curve due to a \ac{PPP} of intensity $\frac{\lambda}{2}$ essentially overlaps with the curve depicting the integration-based results for a lattice with $c\!=\!\lambda^{-1}$ and  $\epsilon\!=\! 0$. Their difference (not possible to notice it in the figure) is the standard deviation of interference due to a lattice without fading. 

It might be useful to derive a closed-form approximation for the difference of the variances for $\epsilon\!>\!0$ and $\epsilon\!=\!0$. We recall it is only the term $J_3$ that depends on $\epsilon$. Therefore we will expand $J_3$ for large $q$ in~\eqref{eq:J23} and~\eqref{eq:J3eps}, and take their difference. With large $q$, the argument of the Zeta function becomes also large, thus it can be well-approximated by an integral instead of a sum. Starting from~\eqref{eq:J23} we get 
\[
\begin{array}{ccl}
J_3 \!\!\!\!&=&\!\!\!\! \displaystyle 2c^{-2\eta} \!\!\!\int_0^1 \!\!\! \zeta\!\left(\eta,q\!+\! x\right) \zeta\!\left(\eta,q\!+\!1\!-\! x\right) \! {\rm d}x \\ 
\!\!\!\!&\approx&\!\!\!\! \displaystyle 2c^{-2\eta}\!\!\!\int_0^1\!\!\!\left(\int_0^\infty \!\!\!\!\!\left(k\!+\!q\!+\!x\right)^{-\eta}\! {\rm d}k \! \int_0^\infty \!\!\!\!\! \left(k\!+\!q\!+\!1\!-\!x\right)^{-\eta}\! {\rm d}k\right) \! {\rm d}x \\ \!\!\!\!&=&\!\!\!\! \displaystyle  \frac{2c^{-2\eta}}{\left(\eta-1\right)^2} \!\int_0^1 \! \left(q\!+\!x\right)^{1-\eta}  \left(q\!+\!1\!-\!x\right)^{1-\eta} {\rm d}x. 
\end{array}
\]

\noindent 
For $q\!\gg\! x$, we may do first-order expansion.
\[
\begin{array}{ccl}
J_3 \!\!\!\!&\approx&\!\!\!\! \displaystyle \frac{2c^{\!-\!2\eta}}{\left(\eta\!-\!1\right)^2} \!\int_0^1 \!\!\! \left(q^{1\!-\!\eta}\!-\!\frac{\left(\eta\!-\! 1\right) x}{ q^\eta}\right) \!\!\! \left(q^{1\!-\!\eta}\!-\!\frac{\left(\eta\!-\! 1\right)\! \left(1\!-\!x\right)}{q^\eta}\right) \!{\rm d}x \\ \!\!\!\!&=&\!\!\!\! \displaystyle  \frac{r_0^{-2\eta} \left(c^2 \left(\eta-1\right)^2 -6c\left(\eta-1\right)r_0 + 6r_0^2 \right)}{3 c^2 \left(\eta-1\right)^2}.
\end{array}
\]

After approximating in a similar manner the term $J_3$ in~\eqref{eq:J3eps} and subtract it from the above, we end up with  $\epsilon\left(\epsilon-1\right)r_0^{-2\eta}$. Therefore the variance of interference due to a lattice of inter-point distance $c$ can be approximated as 
\begin{equation}
\label{eq:VarGrid2}
\mathbb{V}{\text{ar}}\left\{\mathcal{I}\right\} \approx \frac{2 \,  r_0^{1-2\eta}}{c\left(2\eta\!-\!1\right)} \!+\! \epsilon \left(1\!-\!\epsilon\right) r_0^{-2\eta}, 
\end{equation} 
where for $\epsilon\!=\!0$, the variance has been approximated by the variance due to a \ac{PPP} of intensity $\frac{1}{2 c}$. 

The accuracy of~\eqref{eq:VarGrid2} is illustrated in Fig.~\ref{fig:StdGrid}, where it essentially overlaps with the integration-based results. Using $\epsilon\!=\!\frac{1}{2}$ in~\eqref{eq:VarGrid2} indicates that under Rayleigh fading and large cell size, a lattice of intensity $\lambda\!=\!c^{-1}$ can at most increase by $\frac{r_0^{-2\eta}}{4}$  the variance of interference due to a \ac{PPP} of intensity $\frac{\lambda}{2}$. This approximation is also available in  Fig.~\ref{fig:StdGrid}, 'dashed cyan' curve. In Fig.~\ref{fig:StdLam1} and Fig.~\ref{fig:StdLam025}, the selected values of cell size, $r_0$, and intensity $\lambda$ result in $\epsilon\!=\! 0$. We can also observe over there the approximately $\sqrt{2}$-relation of the standard deviations of interference due to a \ac{PPP} and due to a lattice of equal intensity under Rayleigh fading. 

The third moment of interference originated from a lattice can be calculated in a similar manner. The calculation is more cumbersome because triples of sums are involved but it does not come with any new insights. The third-order correlation degenerates to one-dimensional integral with respect to $z$. A low-complexity approximation for the skewness, similar to the one in equation~\eqref{eq:VarGrid2} for the variance, is also possible. 

\section{Conclusions}
\label{sec:Conclusions}
In this paper we have shown that introducing small tracking distance $c$ (as compared to the mean inter-vehicle distance $\lambda^{-1}$) in \ac{1D} vehicular networks reduces the variance of interference exponentially, $e^{-\lambda c}$, with respect to the variance due to a \ac{PPP} of equal intensity $\lambda$. Since the mean interference levels under the two deployment models are equal, the coefficient of variation is reduced by $e^{-\frac{\lambda c}{2}}$, and the interference distribution becomes more concentrated around its mean. Assuming that the hardcore distance is fixed, the exponential correction makes sense to use especially under dense traffic, large $\lambda$. In addition, the distribution of interference for small $\lambda c$ remains positively-skewed, indicating that the gamma distribution would probably provide better fit than the normal distribution. We have also studied the extreme scenario where the interference originates from  infinite \ac{1D} lattice to get some first insight into the properties of interference due to the flow of platoons of vehicles. Under Rayleigh fading and large cell size $r_0$ in comparison with the inter-point lattice distance $c$, we have shown that the term $\left(\frac{2r_0^{1-2\eta}}{c \left(2\eta-1\right)}\!+\! \frac{r_0^{-2\eta}}{4}\right)$ can be used as a tight upper bound for the variance of interference. The results of this paper can serve as a preliminary step before studying the probability of outage in the uplink of vehicular networks with a more realistic deployment model than the \ac{PPP}. Temporal and spatial aspects of interference and more complex headway models are also relevant topics, the reader may refer to~\cite{Koufos2019} for some recent results.

\end{document}